\documentclass[lettersize,journal]{IEEEtran}
\usepackage{amssymb}
\usepackage{amsmath,amsfonts}
\usepackage{algorithmic}
\usepackage{algorithm}
\usepackage{array}
\usepackage[caption=false,font=normalsize,labelfont=sf,textfont=sf]{subfig}
\usepackage{textcomp}
\usepackage{stfloats}
\usepackage{url}
\usepackage{verbatim}
\usepackage{graphicx}
\usepackage{cite}
\usepackage{xcolor}  %
\usepackage{stfloats}
\usepackage{bm}
\usepackage{booktabs}
\makeatletter
\newcommand\multiline[1]{\parbox[t]{\dimexpr\linewidth-\ALG@thistlm}{#1}}
\makeatother
 \DeclareMathSizes{10}{10}{5}{5}
\begin{document}

\title{Direction Modulation Design for UAV Assisted by IRS with discrete phase shift}

\author{Maolin Li, Wei Gao, Qi Wu, Feng Shu, Cunhua Pan, Di Wu 
	\thanks{This work was supported in part by the National Natural Science Foundation of China under Grant U22A2002, Grant 62071234, Grant 62071289, Grant 61972093, and Grant 62301005; in part by Hainan Province Science and Technology Special Fund under Grant ZDKJ2021022; in part by the Scientific Research Fund Project of Hainan University under Grant KYQD(ZR)-21008; in part by the Collaborative Innovation Center of Information Technology, Hainan University, under Grant XTCX2022XXC07. (Corresponding authors: Wei Gao,  Feng Shu, and Qi Wu.) }
	\thanks{Maolin Li, Wei Gao, and Di Wu are with the School
		of Information and Communication Engineering, Hainan University, Haikou,
		570228, China (E-mail: 24110810000008@hainanu.edu.cn, gaowei@epri.sgcc.com.cn, hainuwudi@hainanu.edu.cn).}
		\thanks{Feng Shu is with the School of Information and Communication Engineering and Collaborative Innovation Center of Information Technology, Hainan University, Haikou 570228, China, and also with the School of Electronic and Optical Engineering, Nanjing University of Science and Technology, Nanjing 210094, China (E-mail: shufeng@hainanu.edu.cn).}
		\thanks{Cunhua Pan is with the National Mobile Communications Research Laboratory, Southeast University, China (e-mail:cpan@seu.edu.cn).}
				\thanks{Qi Wu is with the School of Electronic, Information and Electrical Engineering, Shanghai Jiao Tong University, Shanghai 200240, China, and also with the State Key Laboratory of Rail Traffic Control and Safety, Beijing Jiaotong University, Beijing 100044, China (e-mail: wuqi7812@sjtu.edu.cn).}
}

%
%

\maketitle

\begin{abstract}
 As a physical layer security technology, directional modulation (DM) can be combined with intelligent reflect-ing surface (IRS) to improve the security of drone communications. In this paper, a directional modulation scheme assisted by the IRS is proposed to maximize the transmission rate of unmanned aerial vehicle (UAV) secure communication. Specifically, with the assistance of the IRS, the UAV transmits legitimate information and main-tains its constellation pattern at the location of legitimate users on the ground, while the constellation pattern is disrupted at the eavesdropper's location. In order to solve the joint optimization problem of digital weight coefficients, UAV position, and IRS discrete phase shift, firstly, the digital weight vector and UAV position are optimized through power minimization. Secondly, three methods are proposed to optimize IRS phase shift, namely vector trajectory (VT) method, cross entropy vector trajectory (CE-VT) algorithm, and block coordinate descent vector trajectory (BCD-VT) algorithm. Compared to traditional cross entropy (CE) methods and block coordinate descent (BCD) methods, the proposed CE-VT and BCD-VT algorithms can improve transmission rate performance. The numerical results validate the effectiveness of the optimization scheme in IRS assisted UAV communication.
\end{abstract}

\begin{IEEEkeywords}
Intelligent reflecting surface (IRS); directional modulation (DM); unmanned aerial vehicle (UAV); vector trajectory (VT) method; discrete phase shift.
\end{IEEEkeywords}

\section{Introduction}
\IEEEPARstart{U}{Unmanned} aerial vehicles (UAVs) have been extensively researched in wireless communications for their versatility in military, civilian, and scientific applications~\cite{Li2022a,Wang2019,Cao2022}. UAVs can effectively serve as aerial communication relay base stations, significantly enhancing the conditions of long-distance communication channels~\cite{Jiang2023}. With aerial flight capabilities, UAVs can establish short-range line-of-sight (LoS) communication links with various targets, effectively minimizing signal transmission fading. Wireless communications on UAVs face numerous challenges, including maintaining stable communication, ex-tending communication distance, bolstering an-ti-interference abilities, and efficiently managing and utilizing spectrum resources~\cite{Wu2022,Hellaoui2023}. 

Currently, there are many works, which attempt to solve the UAV communication problems. In~\cite{Tarekegn2022}, a novel drone base station control strategy was introduced, leveraging deep reinforcement learning to significantly enhance the transmission coverage and connectivity of wireless communication systems. Further, a comprehensive study was conducted on UAV communications, focusing on the assistance provided by the intelligent reflecting surface (IRS), and the findings revealed that actively altering channels through the deployment of IRS can effectively enhance signal quality~\cite{You2021,Liu2022}. In~\cite{Liu2022a,Zeng2021,Zhang2023}, the method for maximizing system throughput was thoroughly analyzed by concurrently optimizing drone scheduling, trajectory planning, and transmission power allocation, but the optimization of transmission power was not considered. In~\cite{Yu2022}, maximizing energy efficiency was studied by jointly designing power allocation, beamforming, and positioning for UAV. Then, considering collision avoidance and speed constraints, the trajectory optimization problem of UAV was studied~\cite{Liang2022}. 

Further, the importance of UAV communications security has been considered and studied. The high possibility of LoS links and the broadcast nature of air-to-ground UAV communications render it vulnerable to eavesdropping attempts by unauthorized individuals. To address the challenge, two technologies, physical layer security~\cite{Na2022,Wang2020} and secret communication~\cite{Li2021,Wang2022,Su2023,Hu2020}, are mainly used to enhance the security of UAV communications based on upper layer protocol security~\cite{Hassija2021}. In~\cite{Zhou2021}, a comprehensive investigation was conducted on the concealed data collection methods employed by a full-duplex UAV, implementing a continuous hovering design flight trajectory, the detection likelihood was significantly diminished in unscheduled user directions through the utilization of artificial noise.
 
However, achieving concealment on the LoS path poses significant challenges in low-rank channels, as there is a risk of information leakage when the eavesdropping node is situated precisely on the LoS path~\cite{Shu2024}. The research on IRSs provides a solution to this problem. In~\cite{Shi2022}, the secrecy throughput maximization in multi-input multi-output (MIMO) systems was investigated through the deployment of IRS, and experimental results demonstrated that the utilization of these surfaces significantly enhances the confidentiality performance of the system. In~\cite{Li2023,Li2023a}, an element-by-element approach was employed to optimize the discrete phase shift of the IRS, resulting in the design of a secure communication waveform, and the radar-communication integrated waveform was designed at the symbol level, ensuring that the modulated symbols can be accurately decoded only in the designated communication direction~\cite{Li2024}. With the increase of IRS elements, the complexity of this method is high.

Recently, directional modulation (DM), a state-of-the-art physical layer security technology, has demonstrated its ability to facilitate secure communication at the symbol level. can focus the transmitted signal in the desired direction with correct modulation constellation while scrambling the pattern in other directions, which is a more secure technique than simple beamforming~\cite{babakhani08b,daly09a,ding15a}. In~\cite{zhu14a}, a DM design was performed based on a four-dimensional (4-D) antenna array to scramble the signal in undesired directions by time modulation, followed by a time series and static amplitude weighting in~\cite{Chen2021} for the high sidelobe level problem to further enhance security. For the peak-to-average power ratio (PAPR) and nonlinear distortion problems of DM transmitters, two step peak clipping (TPC) and digital predistortion (DPD) schemes were proposed in~\cite{Chen2019}. In~\cite{Nusenu2021}, considering the energy saving security requirements of the Internet of Things, an inverse frequency diversity array was proposed, which adopts a time modulation method. In~\cite{Narbudowicz2022}, an energy-efficient DM scheme was proposed using a uniform circular monopole antenna array, reducing interference to other systems.

\subsection{Major Contributions}
Based on the analysis, UAVs operating as aerial base stations exhibit a high likelihood of line-of-sight (LoS) and possess broadcasting characteristics for communication. When the link distance between UAVs and eavesdroppers (Eves) is shorter than that of legitimate users, or when eavesdroppers are oriented in the same direction as legitimate users, security becomes more vulnerable. To address this challenge, this paper explores the DM design for a UAV communicating with multiple users, aided by the IRS. The main contributions of this paper are summarized as follows.
\begin{itemize}

	\item The IRS-assisted UAV DM design is studied. Due to the serious threat posed by the eavesdropper to legitimate transmissions, we use the IRS to assist UAV in better transmitting confidential information to multiple legitimate users on the ground. Specifically, by optimizing the digital weight vector, UAV position, and IRS phase shift matrix, the received symbols at the eavesdropper position are designed to have low amplitude and disturbed phase, while the received symbols at the user position are designed to satisfy the modulation mode.
	
	\item We derive the power upper bound of uncertain signals and reduce the rate maximization problem to the signal amplitude maximization problem. According to the complementary cumulative distribution function of the standard Gaussian distribution, the signal amplitude under the minimum receiving sensitivity guarantee is derived, and the symbol alignment constraint between the symbol with noise and the desired symbol is transformed into the symbol alignment constraint under the ideal channel, which avoids the estimation of the phase of the noisy signal and improves the robustness of the system.
	
	\item The problem of maximizing transmission rate is studied under the constraints of DM symbol design, receiver sensitivity, maximum transmission power, discrete phase shift, constant modulus, and user position range. For solving the proposed non-convex optimization problem, a UAV position optimization scheme is first designed using the method of minimizing transmission power, and digital weight vectors were obtained through proportional amplification. Given the position of the UAV and the digital weight vector, a low complexity vector trajectory (VT) method is proposed, which is combined with traditional cross entropy (CE) method and block coordinate descending (BCD) method respectively to obtain higher transmission rate performance.
\end{itemize}

\subsection{Organization}
The rest of this article is organized as follows. In Sec. \ref{2}, we present the system model. In Sec. \ref{3}, an optimization method for UAV position and digital weight coefficients is proposed. The IRS phase shift matrix optimization method is presented in Sec. \ref{4}. The simulation results are provided in Sec. \ref{5}, and the conclusion is presented in Sec. \ref{6}.

\subsection{Notation}
Notations: $[\ ]^T$ and $[\ ]^H$ denote transpose and conjugate transpose, respectively; $s$, $\mathbf{s}$, and $\mathbf{S}$ denote scalar, vector, and matrix, respectively; $||\ ||$ denotes the $l_2$ norm; $|\ |$ denotes the absolute value operation; $\text{rank}(\ )$ represents the rank of the matrix; $\otimes$ stands for Kronecker product.

\section{System Model}\label{2}
\begin{figure}
	\centering
	\includegraphics[width=0.49\textwidth, trim = 2 2 2 2,clip]{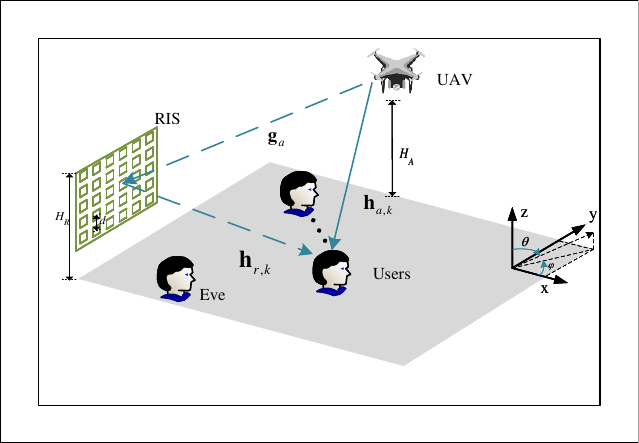}\\
	\caption{System model for UAV-DM assisted by IRS.}\label{fig:1}
\end{figure}
As shown in Fig. \ref{fig:1}, we consider a downlink DM system with $K_u$ ground users, including a single ground point target, i.e., a non-colluded eavesdropper (Eve) with $N_e$ antennas, $n_e = 0,1,\ldots,N_e-1$ represents the $n_e$-th antenna. A UAV equipped with an $N$-element uniform linear array (ULA) as a transmitter and an IRS with $M = M_{Y}M_{Z}$ elements arranged uniformly in a plane is deployed at a height of $H_{R}$, serving $K_u$ single antenna users, where $M_{Y}$ and $M_{Z}$ represent the number of horizontal and vertical elements for the IRS, $M>N$. In three-dimensional space, the depression and azimuth angle are represented by $\theta\in[0,\pi]$ and $\varphi\in[0,\pi]$, respectively. Select the IRS element closest to the UAV as a reference, the coordinate of the UAV is $\mathbf{u}=[\theta_{{A,R}}, \varphi_{{A,R}}, r_{{A,R}}]^{T}$, where $\theta_{{A,R}}$, $\varphi_{{A,R}}$, and $r_{{A,R}}=H_{u} /\cos\theta_{{A,R}}$ are the depression angle, azimuth angle, and distance from the UAV to the IRS, respectively. $H_u$ is the height of the UAV from the ground. We represent the user set as $\mathcal K = \{0,1,...,K_{u}-1\}$, $K_{u}<N<M$. User $k$ and Eve $e$ are located near the IRS, with coordinates represented as  $\mathbf{q}_{k}=[\theta_{{R,k}},\varphi_{{R,k}},r_{{R,k}}]^{T}$ and $\mathbf{v}_{e}=[\theta_{{R,e}},\varphi_{{R,e}},r_{{R,e}}]^{T}$, where $\theta_{{R,k}}$ ($\theta_{{R,e}}$), $\varphi_{{R,k}}$($\varphi_{{R,e}}$), and $r_{{R,k}}=-H_{R} /\cos\theta_{{R,k}}$ ($r_{{R,e}}$) are the depression angle, azimuth angle, and distance from the IRS to the user (Eve), respectively. The IRS can be deployed to maintain communication with users in the near-field area, and the line-of-sight (LoS) path from the UAV to ground users is typically considered far-field communication, considering situations where users are in a mixed near-field and far-field propagation environment. Considering that the maximum phase difference between the spherical wave and plane wave model is no more than $pi/8$, the boundary between the near-field and far-field, i.e., the Fraunhofer distance or Rayleigh distance, is expressed as $2D^2/\lambda$, where $D$ represents the aperture of the antenna, and $\lambda$ represents the wavelength of the carrier \cite{Cui2022}.

In DM, $B$ signaling symbols are designed to fall exactly in the demodulation area of the receiving constellation point at the user's location, avoiding channel estimation at the user's location and increasing additional transmission power to disrupt the receiving constellation pattern at the eavesdropper's location. Assuming that the channels from the UAV to the IRS and from the IRS to the user $k$ (Eve $e$) are represented as $\mathbf{G}\in\mathbb{C}^{M\times N}$, $\mathbf{h}_{{R,k}}\in\mathbb{C}^{M\times 1} (\mathbf{H}_{{R,e}}\in\mathbb{C}^{M\times N_e})$, and $\mathbf{h}_{{A,k}}\in\mathbb{C}^{N\times 1}\left(\mathbf{H}_{{A,e}}\in\mathbb{C}^{N\times N_e}\right)$, respectively, and are defined as UAV-IRS channel, IRS-user (IRS-Eve) channel, and UAV-User (UAV-Eve) channel. The $b$-th$ (b = 0,1,...,B-1)$ symbol received by the $k$-th user is given as
\begin{equation}
	\label{eq1}
	y_{{b,k}}=(\mathbf{h}_{{R,k}}^{T}\mathbf{\Phi}\mathbf{G}+\mathbf{h}_{{A,k}}^{T})\sum_{i=0}^{K_{u}-1}\mathbf{v}_{b,i}s_{b,i}+n_{b,k},
\end{equation}
where $\mathbf{\Phi}=$diag$(\gamma_{0}e^{j\phi_0},\gamma_{1}e^{j\phi_1},...,\gamma_{{M-1}}e^{j\phi_{M-1}})\in\mathbb{C}^{M\times M}$ represents the reflection coefficient matrix at the IRS, $\mathbf{v}_{b,i}\in\mathbb{C}^{N\times1}$ represents the beamforming vector corresponding to the $b$-th unit energy symbol $s_{b,i}$, $n_{b,k}\sim\mathbb{C}\mathbb{N}(0,\sigma^2)$ represents the noise at the user assuming that the noise at all user receiving ends is independent and identically distributed. Consider a passive IRS, where the adjustable amplitude and phase of the $m$-th $( m= 0, 1,...,M- 1)$ reflective unit satisfy $\gamma_{m}=1$ and $\phi_{m}\in[0,2\pi]$. The symbol received by Eve's $n_e$-th antenna is disturbed, which can be expressed as
\begin{equation}
	\label{eq2}
y_{{b,n_e}}=(\mathbf{h}_{{R,n_e}}^{T}\mathbf{\Phi}\mathbf{G}+\mathbf{h}_{{A,n_e}}^{T})\sum_{i=0}^{K_{u}-1}\mathbf{v}_{b,i}s_{b,i}+n_{b,n_e},\end{equation}
where $n_{b,e} \sim \mathbb{CN}( 0, \sigma ^2)$ represents the noise at the Eve, $\mathbf{h}_{{A,n_e}}$ and $\mathbf{h}_{{R,n_e}}$ are respectively  the UAV-Eve channel and the IRS-Eve channel corresponding to the $n_e$-th antenna. Note that to enhance confidentiality performance, by optimizing $\mathbf{w}_{b}$ and $\mathbf{\Phi}$, the amplitude of the phase random complex signal $y_{{b,n_e}}$ is smaller than $y_{{b,k}}$.

Multi-user interference in multi-user scenarios may reduce the transmission rate. To eliminate interference from other users, zero forcing (ZF) method can be used. Then, only the multi-user interference of user $k$ is eliminated, while the eavesdropper $e$ suffers from multi-user interference, we have
\begin{equation}\begin{split}
		\label{eq14-5}
		&(\mathbf{h}_{{R,k}}^{T}\mathbf{\Phi}\mathbf{G}+\mathbf{h}_{{A,k}}^{T})\sum_{i=0,i\neq k}^{K_{u}-1}\mathbf{v}_{b,i}s_{b,i}=0,\\
		&(\mathbf{h}_{{R,n_e}}^{T}\mathbf{\Phi}\mathbf{G}+\mathbf{h}_{{A,n_e}}^{T})\sum_{i=0,i\neq e}^{K_{u}-1}\mathbf{v}_{b,i}s_{b,i}\neq0,\end{split}
\end{equation}
Therefore, user $k$ can correctly demodulate and obtain the transmitted symbols, i,e., 
\begin{equation}\begin{split}
		\label{eq14-6}
		\hat y_{{b,k}}&=\mathbf{v}_{k}^H(\mathbf{h}_{{R,k}}^{T}\mathbf{\Phi}\mathbf{G}+\mathbf{h}_{{A,k}}^{T})^H(\mathbf{h}_{{R,k}}^{T}\mathbf{\Phi}\mathbf{G}+\mathbf{h}_{{A,k}}^{T})\mathbf{v}_{k}s_{b,k}+n_{b,k}\\&=t_{b,k}s_{b,k}+n_{b,k},
	\end{split}
\end{equation}
where $t_{b,k}$ is the amplitude of the symbol received by user $k$. 

For Eve $e$, if the channel information of the user is known, equipping more antennas than the number of users of a single antenna can correctly demodulate the transmitted symbols. Therefore, this information transmission method is not secure. An effective method is to add artificial noise (AN) $z_{b}$ to the baseband signal, and the symbols received by eavesdropper $e$ and user $k$ can be represented as \cite{hu17a, Shu2021}
\begin{equation}\begin{split}
		\label{eqnew5}\begin {cases}
		&\tilde y_{{b,k}}=(\mathbf{h}_{{R,k}}^{T}\mathbf{\Phi}\mathbf{G}+\mathbf{h}_{{A,k}}^{T})\mathbf{v}_{b,k}(s_{b,k}+z_b)+n_{b,k},\\&\tilde y_{{b,n_e}}=(\mathbf{h}_{{R,n_e}}^{T}\mathbf{\Phi}\mathbf{G}+\mathbf{h}_{{A,n_e}}^{T})\sum_{i=0}^{K_{u}-1}\mathbf{v}_{b,i}(s_{b,i}+z_{b})+n_{b,n_e},\end {cases}
	\end{split}
\end{equation}
Due to $z_{b}$ being designed in the null space of user channel, i.e., $(\mathbf{h}_{{R,k}}^{T}\mathbf{\Phi}\mathbf{G}+\mathbf{h}_{{A,k}}^{T})\mathbf{v}_{k}z_{b}=0$, user $k$ can correctly demodulate symbols. 

However, the traditional approach of adding AN to the baseband requires channel estimation at the receiver. One way for a receiver to not require channel estimation is through transmitter precoding, which synthesizes symbols at the user and designs difficult to demodulate signal patterns at the eavesdropper, i.e., the user does not need $\mathbf{v}_{k}^H(\mathbf{h}_{{R,k}}^{T}\mathbf{\Phi}\mathbf{G}+\mathbf{h}_{{A,k}}^{T})^H$ to demodulate the signal, i.e.,
the receive symbol $\hat y_{b,k}^*$ at the user is
\begin{equation}
	\label{eqnew6} 
	\hat y_{b,k}^*=t_{b,k}s_{b,k}+n_{b,k}.
\end{equation}

Then, by introducing $\mathbf{w}$, the baseband signal is designed and can be written as
\begin{equation}
	\label{eq14-3}
	\mathbf{w}_b=\sum_{i=0}^{K_{u}-1}\mathbf{v}_{i}(s_{b,i}+z_{b}).
\end{equation}
Correspondingly, the signals received by Eve $e$ and user $k$ can be represented as
\begin{equation}
	\label{eqnew7}\begin{cases}
	y_{{b,k}}=(\mathbf{h}_{{R,k}}^{T}\mathbf{\Phi}\mathbf{G}+\mathbf{h}_{{A,k}}^{T})\mathbf{w}_b+n_{b,k},\\
	y_{{b,n_e}}=(\mathbf{h}_{{R,n_e}}^{T}\mathbf{\Phi}\mathbf{G}+\mathbf{h}_{{A,n_e}}^{T})\mathbf{w}_b+n_{b,n_e}.
	\end{cases}
\end{equation}

\subsection{Channel Model}
Considering that the UAV and the IRS are typically deployed at a certain height from ground users, the actual UAV-IRS, IRS-user, and UAV-user (UAV-Eve) channels consist of LoS-dominated components and low-power non-line-of-sight (NLoS) components. Assuming parameter $g\in\{\mathcal K,e\}$ represents all receivers, i.e., $g\triangleq k$ represents the user and $g\triangleq e$ denotes the eavesdropper. Then, $\mathbf{G}$, $\mathbf{h}_{{R,g}}$, and $\mathbf{h}_{{A,g}}$ can be modelled as
\begin{equation}
	\begin{split}
	\label{eq3}
	\mathbf{G}&=\sqrt{\vphantom{()}\alpha_{{A,R}}\varepsilon_{{A,R}}}\overline{\mathbf{G}}+\sqrt{\alpha_{{A,R}}(1-\varepsilon_{{A,R}})}\hat{\mathbf{G}},\\
	\mathbf{h}_{{R,g}}&=\sqrt{\vphantom{()}\alpha_{{R,g}}\varepsilon_{{R,g}}}\overline{\mathbf{h}}_{{R,g}}+\sqrt{\alpha_{{R,g}}(1-\varepsilon_{{R,g}})}\hat{\mathbf{h}}_{{R,g}},\\
	\mathbf{h}_{{A,g}}&=\sqrt{\vphantom{()}\alpha_{{A,g}}\varepsilon_{{A,g}}}\overline{\mathbf{h}}_{{A,g}}+\sqrt{\alpha_{{A,g}}(1-\varepsilon_{{A,g}})}\hat{\mathbf{h}}_{{A,g}},
	\end{split}
\end{equation}
where $\alpha_{{A,R}}=\rho/d_{{A,R}}^{2}$, $\alpha_{{R,g}}=\rho/d_{{R,g}}$ and $\alpha_{{A, g}}=\rho/d_{{A,g}}^{2}$represent the path loss coefficients in free space, $\rho$ is the channel power gain per unit distance, $d_{{A, R}}=\|\mathbf{u}\|$, $d_{{R, k}}= \|\mathbf{q}_{k}\|$ $(d_{{R, e}}=\|\mathbf{v}_{e}\|)$, and $d_{{A,k}}=\|\mathbf{q}_{k}-\mathbf{u}\| (d_{{A,e}}=\|\mathbf{v}_{e}-\mathbf{u}\|)$ represent the distances from the UAV to the IRS, the IRS to the user $k$ (Eve $e$) and the UAV to the user $k$ (Eve $e$), respectively. $\varepsilon_{{A, R}}$, $\varepsilon_{{R, g}}$ and $\varepsilon_{A,g}$ represent the LoS power ratio coefficients of the corresponding channels. $\bar{\mathbf{G}},\bar{\mathbf{h}}_{{R,g}}$, and $\bar{\mathbf{h}}_{{A,g}}$ represent the LoS component. respectively, given as
\begin{equation}
	\begin{split}
		\label{eq4}
	\mathbf{\bar{G}}&=(\mathbf{a}_{{A,R}}\otimes\mathbf{b}_{{A,R}})\mathbf{h}_{{A,R}}^{T}, \\
	{\overline{\mathbf{h}}}_{{R,g}}&=\mathbf{a}_{{R,g}}\otimes\mathbf{b}_{{R,g}},\\
	{\overline{\mathbf{h}}}_{{A,g}}(\theta_{{A,g}}) &= \left[1,e^{j\frac{2\pi d_{A}\cos\theta_{{A,g}}}{\lambda}},...,e^{j\frac{2\pi(N-1)d_{A}\cos\theta_{{A,g}}}{\lambda}} \right]^{T}, \\
	\end{split}
\end{equation}
where $d_A$ is the minimum spacing between array elements of the UAV antenna array, $\lambda$ is the wavelength, $\mathbf{h}_{{A,R}}\triangleq\mathbf{h}_{{A,R}}(\theta_{{A,R}})\in\mathbb{C}^{N\times1}$, $\mathbf{b}_{{A,R}}\triangleq\mathbf{b}_{{A,R}}(\varphi_{{A,R}},\theta_{{A,R}})\in\mathbb{C}^{M_{Y}\times1}$, and $\mathbf{a}_{{A,R}}\triangleq\mathbf{a}_{{A,R}}(\theta_{{A,R}})\in\mathbb{C}^{M_{Z}\times1}$ represent the steering vectors from the UAV to the reference unit of the IRS, from the UAV to the vertical and horizontal dimensions of the IRS, respectively, are given as
\begin{equation}
	\begin{split}
		\label{eq5}
	\mathbf{h}_{{A,R}}& =\left[1,e^{j\frac{2\pi d_{A}\cos\theta_{{A,R}}}{\lambda}},...,e^{j\frac{2\pi(N-1)d_{A}\cos\theta_{{A,R}}}{\lambda}}\right]^{T}, \\
	\mathbf{a}_{{A,R}}&= \left[1,e^{j\frac{2\pi d_{R}\cos\theta_{{A,R}}}{\lambda}},...,e^{j\frac{2\pi d_{R}(M_Y-1)\cos\theta_{{A,R}}}{\lambda}}\right]^{T}, \\
	\mathbf{b}_{{A,R}}&=\biggl[1,e^{j\frac{2\pi d_{R}\sin\varphi_{{A,R}}\sin\theta_{{A,R}}}{\lambda}},
	\\&...,e^{j\frac{2\pi(M_Z-1)d_{R}\sin\varphi_{{A, R}}\sin\theta_{{A, R}}}{\lambda}} \Big\rceil^{T},
	\end{split}
\end{equation}
$\mathbf{a}_{{R,g}}\triangleq\mathbf{a}_{{R,g}}(r_{{\nu,g}})\in\mathbb{C}^{M_{Z}\times1}$ and $\mathbf{b}_{{R,g}}\triangleq\mathbf{b}_{{R,g}}(r_{{h,g}})\in\mathbb{C}^{M_{Y}\times1}$ represent the vertical and horizontal steering vectors from the IRS to the receiver, respectively, written as
\begin{equation}
	\begin{split}
			\label{eq6}
	\mathbf{a}_{{R,g}}(r_{{\nu,g}})&=e^{j2\pi r_{g}/\lambda}\left[1,e^{-j2\pi r_{{1,g}}/\lambda},...,e^{j2\pi r_{{M_{Z}-1,g}}/\lambda}\right]^T,\\
	\mathbf{b}_{{R,g}}(r_{{h,g}})&=e^{j2\pi r_g/\lambda}\left[1,e^{-j2\pi r_{{1,g}}/\lambda},...,e^{j2\pi r_{{M_{Y}-1,g}}/\lambda}\right]^T.	
	\end{split}
\end{equation}
where $d_{R}$ is the minimum spacing between the IRS elements, $r_{g}$ is the distance from the IRS reference unit to the receiver. According to geometric relationships, the distances $r_{{\nu,g}}$ and $r_{{h,g}}$ from the $\nu$-th $(\nu=0,1,...,M_{Z}-1)$ individual element in the vertical direction and the $h$-th $(h=0,1,...,M_{Y}-1)$ element in the horizontal direction of the IRS to the receiver corresponding to the receiver can be given as
\begin{equation}
	\begin{split}
		\label{eq7}
r_{{\nu,g}}&=\sqrt{r_{{R,g}}^{2}+\nu^{2}d^{2}-2r_{{R,g}}\nu d\cos\theta_{{R,g}}},\\
r_{{h,g}}&=\sqrt{r_{{R,g}}^{2}+h^{2}d^{2}-2r_{{R,g}}hd\sin\varphi_{{R,g}}\sin\theta_{{R,g}}},
	\end{split}
\end{equation}
respectively. $\hat{\mathbf{G}}$, $\hat{\mathbf{h}}_{{R,g}}$ and $\hat{\mathbf{h}}_{{R,g}}$ are an independent and identically distributed complex Gaussian random variable with a mean of zero and unit covariance, representing the NLoS component. Accordingly, the aggregated channel from the UAV to the receiver can be represented as a combination of LoS and NLoS components, i.e.,
\begin{equation}
\begin{split}
	\label{eq8}
	\mathbf{h}_{\mathrm{g}}& =\mathbf{h}_{{R,g}}^T\mathbf{\Phi}\mathbf{G}+\mathbf{h}_{{A,g}}^{T} \\
	&=(\underbrace{\sqrt{\vphantom{()}\alpha_{{R,g}}\varepsilon_{{R,g}}}}_{{\zeta_{g}^{1}}}\overline{{\mathbf{h}}}_{{R,g}}^{T}+\underbrace{\sqrt{\vphantom{()}\alpha_{{R,g}}(1-\varepsilon_{{R,g}})}}_{{\zeta_{g}^{2}}}\hat{\mathbf{h}}_{{R,g}}^{T})\boldsymbol{\Phi}\\
	&\quad\  (\underbrace{\sqrt{\vphantom{()}\alpha_{{A,R}}\varepsilon_{{A,R}}}}_{{\upsilon^{1}}}\overline{{\mathbf{G}}}+\underbrace{\sqrt{\alpha_{{A,R}}(1-\varepsilon_{{A,R}})}}_{{\upsilon^{2}}}\hat{\mathbf{G}})\\
	&+\underbrace{\sqrt{\vphantom{()}\alpha_{A,g}\varepsilon_{A,g}}}_{{\zeta_{g}^{3}}}\overline{{\mathbf{h}}}_{{A,g}}^{T}+\underbrace{\sqrt{\alpha_{{A,g}}(1-\varepsilon_{{A,g}})}}_{{\zeta_{g}^{4}}}\hat{\mathbf{h}}_{A,g}^{T} \\
	&=\underbrace{\zeta_{g}^{1}\upsilon^{1}\overline{{\mathbf{h}}}_{{R,g}}^{T}\mathbf{\Phi}{\overline{{\mathbf{G}}}+\zeta_{g}^{3}\overline{{\mathbf{h}}}_{A,g}^{T}}}_{{\mathbf{h}_{g}^{LoS}}}\\
	&+\underbrace{{\zeta_{g}^{2}\upsilon^{1}\hat{\mathbf{h}}_{{R,g}}^{T}\mathbf{\Phi}\overline{{\mathbf{G}}}+\zeta_{g}^{1}\upsilon^{2}\overline{{\mathbf{h}}}_{{R,g}}^{T}\mathbf{\Phi}\hat{\mathbf{G}}}}_{{\mathbf{h}_{g}^{{NLoS}}}}\\
	&+\underbrace{{\zeta_{g}^{2}\upsilon^{2}\hat{\mathbf{h}}_{{R,g}}^{T}\mathbf{\Phi}\hat{\mathbf{G}}+\zeta_{g}^{4}\hat{\mathbf{h}}_{{A,g}}^{T}}}_{{\mathbf{h}_{g}^{{NLoS}}}} \\
	&=\mathbf{h}_{g}^{LoS}+\mathbf{h}_{g}^{NLoS}.
\end{split}
\end{equation}
Then, the $b$-th symbol received by the receiver can be rewritten as
\begin{equation}
	\begin{split}
		\label{eq9}
	y_{{b,g}}& =(\mathbf{h}_{g}^{LoS}+\mathbf{h}_{g}^{NLoS})\mathbf{w}_{b}+n_{g} \\
	&=\mathbf{h}_{g}^{LoS}\mathbf{w}_{b}+\hat{n}_{{b,g}},
\end{split}
\end{equation}
where $\hat{n}_{{b,g}}=\mathbf{h}_{g}^{NLoS}\mathbf{w}_{b}+n_{g}$ is the sum of all uncertain terms, following a Gaussian distribution with modified variance, i.e., $\hat{n}_{b,g}\sim\mathbb{CN}(0,\hat{\sigma}_{b,g}^2)$, satisfying
\begin{equation}
	\begin{split}
		\label{eq10}
	\hat{\sigma}_{b,g}^{2}& =(\parallel\zeta_g^2\upsilon^1\overline{\mathbf{G}}\boldsymbol{\Phi}\mathbf{w}_b\parallel^2+\frac12\parallel\zeta_g^2\upsilon^2\boldsymbol{\Phi}\otimes\mathbf{w}_b\parallel_F^2 \\
	&+(\zeta_g^4)^2\parallel\mathbf{w}_b\parallel^2+\parallel\zeta_g^1\upsilon^2(vec(\mathbf{\Phi})\cdot\overline{\mathbf{h}}_{R,g}^{T})\otimes\mathbf{w}_b\parallel_F^2+\sigma^2) \\
	&= \parallel\zeta_g^2\upsilon^1\mathbf{\bar{G}}\mathbf{w}_b\parallel^2 +(\frac12M(\zeta_g^2\upsilon^2)^2\\
	&+(\zeta_g^4)^2+M(\zeta_g^1\upsilon^2)^2)\|\mathbf{w}_b\|^2+\sigma^2.
	\end{split}
\end{equation}
According to \eqref{eq9} and \eqref{eq10}, for the $b$-th symbol, the signal-to-noise ratio (SNR) at the receiver can be written as
\begin{equation}
	\begin{split}
		\label{eq11}
		\text{SNR}_{b,g}=\frac{\parallel\mathbf{w}_b\mathbf{h}_g^{LoS}\parallel^2}{\hat{\sigma}_{b,g}^2}=\frac{t_{b,g}^2}{\hat{\sigma}_{b,g}^2},
	\end{split}
\end{equation}
where $t_{b,g}$ is the magnitude corresponding to the $b$-th symbol.

For an ideal beamforming where the beam gain in an undesired direction is zero, the sum of the beam gains in all desired directions is equal to the number of transmitting antennas. In practice, it is difficult to achieve ideal beamforming, and the sum of beam gains satisfies		
\begin{equation}
	\begin{split}
		\label{eq12}
\parallel\bar{\mathbf{G}}\mathbf{w}_b\parallel^2\leq N.
	\end{split}
\end{equation}
According to \eqref{eq10} and \eqref{eq12}, the upper bound on the variance of the sum of uncertain terms can be obtained, i.e.,
\begin{equation}
	\begin{split}
		\label{eq13}
		\hat{\sigma}^{2}_{b,g}&\leq( (\zeta_g^2\upsilon^1)^2\frac NM+\frac12(\zeta_g^2\upsilon^2)^2+\frac{(\zeta_g^4)^2}M+(\zeta_g^1\upsilon^2)^2)MP_{\max}+\sigma_g^2\\&=\hat{z}_g,
	\end{split}
\end{equation}
where $P_{max}$ is the maximum transmit power threshold for the UAV. Then, the lower bound of the transmission rate of the user $k$ is expressed as 
\begin{equation}
	\begin{split}
		\label{eq14}
R_{b,g}=\log(1+t_{b,g}^2/\hat{z}_g), g\triangleq k.
	\end{split}
\end{equation}

\subsection{DM design assisted by IRS}
Let $t_{b,k}$ and $t_{b,e}$ represent the amplitude of the received symbols at user $k$ and Eve $e$, $\vartheta_{b,k}$ and $\vartheta_{b,e}$ represent the phase of the received symbols at the user and eavesdropper, respectively. The desired constellation pattern $\{t_{b,k}e^{j\vartheta_{b,k}}\}_{b=0}^{B-1}$ is synthesized in user directions, while is disrupted to $\{t_{b,e}e^{j\vartheta_{b,e}}\}_{b=0}^{B-1}$ in Eve directions. With $\hat{z}_g$ obtained by \eqref{eq13}, maximizing the transmission rate at the user is equivalent to maximizing $t_{b,k}$ according to monotonicity. Then, in the constellation pattern design of users and the Eve, the problem of maximizing the user transmission rate under the constraints of the minimum detectable symbol power, maximum transmit power, constant reflection amplitude, discrete phase shift, and positioning range can be formulated as
\begin{equation}
	\begin{split}
		\label{eq15}
	\begin{aligned}
		P1:&\max_{\mathbf{w}_{b}, \mathbf{\Phi}, \mathbf{u}}\sum_{k=0}^{K_{u}-1}t_{b,k} \\
		\mathrm{s.t.}\  &\text{Cl} :\mathbf{h}_{k}^{LoS}\mathbf{w}_{b}+\hat{n}_{b,k}=t_{b,k}e^{j\vartheta_{b,{k}}},\forall k, \\
		&\text{C2} :\mathbf{h}_{e}^{LoS}\mathbf{w}_{b}+\hat{n}_{b,e}=t_{b,e}e^{j\vartheta_{b,e}},\forall e, \\
		&\text{C3} :t_{b,k}\geq r_{\mathrm{min}},\forall k,  \\
		&\text{C4} :\parallel\mathbf{w}_{b}\parallel^{2}\leq P_{\max}, \\
		&\text{C5} :\gamma_{m}=1,\phi_{m}\in[0,2\pi],\forall m, \\
		&\text{C6} :\theta_{\min}\leq\theta_{A,R}\leq\theta_{\max},\varphi_{\min}\leq\varphi_{A,R}\leq\varphi_{\max}, 
	\end{aligned}
	\end{split}
\end{equation}
where, $r_\mathrm{min}$ is the minimum received symbol power threshold for the user, reflecting the sensitivity of the user's received signal, i.e., the minimum symbol power that the user can detect is $r_\mathrm{min}.$ $\theta _\mathrm{min}$ and $\theta_\mathrm{max}$ represent the minimum and maximum elevation angles of the UAV, $\varphi_\mathrm{min}$ and $\varphi_\mathrm{max}$ represent the minimum and maximum azimuth angles of the UAV, respectively.

However, for the equality constraints $C1$ and $C2$, the received symbol phases are strictly aligned with the desired phases, and the phase of uncertain items $\hat{n}_{b,k}$ and $\hat{n}_{b,e}$ are difficult to obtain accurately. The derivation for designing under the presumption of a noise-free environment is as follows. 

Based on the concept of constructive interference~\cite{Wei2020}, the symbols received at the user $k$ can be designed by relaxing the phase, which is written as
\begin{equation}
	\begin{split}
		\label{eq16}
\begin{aligned}&\mid\mathfrak{S}\{(\mathbf{h}_{k}^{LoS}\mathbf{w}_{b}+\hat{n}_{b,k})e^{j\vartheta_{b,k}}\}|\leq\\&(\Re\{(\mathbf{h}_{k}^{LoS}\mathbf{w}_{b}+\hat{n}_{b,k})e^{j\vartheta_{b,k}}\}-t_{b,k})\tan\varpi,\forall k,b,\end{aligned}
	\end{split}
\end{equation}
where $\varpi=\pi/B$ represents the maximum phase deviation from the standard constellation point. However, due to receiving noise and power, \eqref{eq16} may not always be satisfied, resulting in erroneous decoding. To reduce the bit error rate (BER), based on the linear property of Gaussian distribution, \eqref{eq16} can be written as \eqref{eq17}, as shown at the top of the next page, where $\Gamma$ is the detection probability threshold, which is the complementary cumulative distribution function with variance $(1+\tan^2\varpi)\hat{\sigma}_{b,k}^2.$ The inverse cumulative distribution function of the standard Gaussian variable is represented as $\Phi^{-1}(\cdot)$, \eqref{eq16} can be written as
\begin{figure*}[ht]
	\centering
	\begin{eqnarray}
		\label{eq17}
		\Pr\{\frac{\mid\mathfrak{T}\{\hat{n}_{b,k}e^{{j\vartheta_{b,k}}}\}\mid-\Re\{\hat{n}_{b,k}e^{{j\vartheta_{b,t}}}\}\tan\varpi}{\sqrt{(1+\tan^{2}\varpi)\hat{\sigma}_{b,k}^{2}}}>\frac{(\Re\{\sqrt{p}\mathbf{h}_{k}^{LoS}\mathbf{w}_{b}e^{{j\vartheta_{b,k}}}\}-t_{{_{u}}})\tan\varpi-\mid\mathfrak{T}\{\sqrt{p}\mathbf{h}_{k}^{LoS}\mathbf{w}_{{{b}}}e^{{j\vartheta_{b,k}}}\}\mid}{\sqrt{(1+\tan^{2}\varpi)\hat{\sigma}_{b,k}^{2}}}\}<\Gamma,\forall k,b.	
	\end{eqnarray}
		\vspace*{8pt}
	\hrulefill
	\vspace*{8pt} 
\end{figure*}
\begin{equation}
	\begin{split}
		\label{eq18}
\begin{aligned}&\mid\mathfrak{J}\{\mathbf{h}_{k}^{LoS}\mathbf{w}_{b}e^{j\vartheta_{b,k}}\}\mid\leq\\&(\Re\{\mathbf{h}_{k}^{LoS}\mathbf{w}_{b}e^{j\vartheta_{b,k}}\}-t_{b,k})\tan\varpi-\Delta_{k},\forall k,b,\end{aligned}
	\end{split}
\end{equation}
where $\Delta_k=\Phi^{-1}(1-\Gamma)\sqrt{(1+\tan^2\varpi)z_k}.$ The right-hand term of \eqref{eq18} is greater than or equal to zero, and the larger its value, the better the receiver's reception performance. With $(\Re\{\mathbf{h}_{k}^{LoS}\mathbf{w}_{b}e^{j\vartheta_{b,k}}\}-t_{b,k})\tan\varpi-\Delta_{k}=0$, estimate the uncertain item $\hat{n}_{b,g}$ and ensure that the symbol amplitude $\hat{t}_{b,k}\geq r_{\mathrm{min}}+\Delta_{k}/\tan\varpi=r_{min,k}$ to meet the minimum signal reception power at the user's location. Then, we have $\Re\{\mathbf{h}_{k}^{LoS}\mathbf{w}_{b}e^{j\vartheta_{b,k}}\}=\hat{t}_{b,k}$. For eavesdropping locations, low amplitude and phase random symbols are designed. Therefore, the constraints $C1$ and $C2$ can be redescribed as 
\begin{equation}
	\begin{split}
		\label{eq20_1}
		\begin{aligned}
&\text{C1}^{\prime} :\mathbf{h}_{k}^{LoS}\mathbf{w}_{b}=\hat{t}_{b,k}e^{j\vartheta_{b,{k}}},\forall k, \\
			&\text{C2}^{\prime}: \mathbf{h}_{e}^{LoS}\mathbf{w}_{b}=t_{b,e}e^{j\vartheta_{b,e}},\forall e, \\
		\end{aligned}
	\end{split}
\end{equation}
respectively.

For the phase shift constraint $C5$ of IRS, in practice, the accuracy of phase shift is limited, i.e., $\phi_m$ can only be selected from a finite set $\mathbb{F}$, given by
\begin{equation}
	\begin{split}
		\label{eq19}
\phi_m\in\mathbb{F}=[0,2\pi/2^{\tilde{B}},...,(2^{\tilde{B}}-1)2\pi/2^{\tilde{B}}],
	\end{split}
\end{equation}
where $\tilde{B}$ represents $\tilde{B}$ bits. Then, the problem of maximizing user transmission rate is formulated as
\begin{equation}
	\begin{split}
		\label{eq20}
\begin{aligned}
	P2:&\max_{\mathbf{w}, \mathbf{\Phi}, \mathbf{u}} \sum_{k=0}^{K_u-1}\hat{t}_{b,k} \\
	\mathrm{s.t.}\ &\text{C1}^{\prime} :\mathbf{h}_{k}^{LoS}\mathbf{w}_{b}=\hat{t}_{b,k}e^{j\vartheta_{b,{k}}},\forall k, \\
	&\text{C2}^{\prime}: \mathbf{h}_{e}^{LoS}\mathbf{w}_{b}=t_{b,e}e^{j\vartheta_{b,e}},\forall e, \\
	&\text{C3} ^{\prime}:\hat{t}_{k}\geq r_{min,k},\\
	&\text{C4} :\parallel\mathbf{w}_b\parallel^2\leq P_{\max} , \\
	&\text{C5}^{\prime} :\gamma_m=1,\phi_m\in\mathbb{F},\forall m, \\
	&\text{C6}:\theta_{\mathrm{min}}\leq\theta_{A,R}\leq\theta_{\mathrm{max}},\varphi_{\mathrm{min}}\leq\varphi_{A,R}\leq\varphi_{\mathrm{max}}. 
\end{aligned}
	\end{split}
\end{equation}
However, the constraints $C5$ and $C$6 are non-convex, and the optimization variables $\mathbf{w}_{b}$ and $\mathbf{\Phi}$ are coupled with each other. For the case of a large number of the IRS elements, the highly complex exhaustive search method is not applicable.

\subsection{DoF analysis}
For the far-field assumption, the degree of freedom (DoF) $\text{DoF}_e$ of MIMO communication without IRS deployment at Eve can be expressed as
\begin{equation}
	\begin{split}
		\label{eq14-1}
		\text{DoF}_e\leq\min(N,N_e).
	\end{split}
\end{equation}
For $K_u$ single-antenna multiple access users, the maximum degree of freedom is $K_u$ \cite{Tse2005}.

Furthermore, deploying MIMO communication with a IRS, the degree of freedom $\text{DoF}_{e,0}$ at Eve can be represented as
\begin{equation}
	\begin{split}
		\label{eq14-2}
		\text{DoF}_{e,0}&\leq\text{rank}(\mathbf{H}_{{R,e}}\mathbf{\Phi}\bar{\mathbf{G}}+\mathbf{H}_{{A,e}})\\&\leq \text{rank}(\mathbf{H}_{{R,e}}\mathbf{\Phi}\bar{\mathbf{G}})+\text{rank}(\mathbf{H}_{{A,e}})\\
		&\leq \min(\text{rank}(\mathbf{H}_{{R,e}}\mathbf{\Phi}),\text{rank}(\bar{\mathbf{G}}))+\text{rank}(\mathbf{H}_{{A,e}})\\
		&\leq \min(\min(\text{rank}(\mathbf{H}_{{R,e}}),M),\text{rank}(\bar{\mathbf{G}}))+\text{rank}(\mathbf{H}_{{A,e}})\\&=\min(\text{rank}(\mathbf{H}_{{R,e}}),\text{rank}(\bar{\mathbf{G}}))+\text{rank}(\mathbf{H}_{{A,e}}),
	\end{split}
\end{equation}
In LOS channel, $\text{rank}({\mathbf{H}}_{R,e})=\text{rank}(\bar{\mathbf{G}})=\text{rank}({\mathbf{H}}_{A,e})=1$, so we have
\begin{equation}
	\begin{split}
		\label{eq14-2-1}
		\text{DoF}_{e,0}&\leq2.
	\end{split}
\end{equation}
the degree of freedom $\text{DoF}_{e,0}$ at Eve can be represented as
\begin{equation}
	\begin{split}
		\label{eq14-2-2}
		\text{DoF}_{u,0}&\leq\text{rank}(\mathbf{H}_{{R,K_u}}\mathbf{\Phi}\bar{\mathbf{G}}+\mathbf{H}_{{A,K_u}})\\&\leq \text{rank}(\mathbf{H}_{{R,K_u}}\mathbf{\Phi}\bar{\mathbf{G}})+\text{rank}(\mathbf{H}_{{A,K_u}})\\
		&\leq \min(\text{rank}(\mathbf{H}_{{R,K_u}}\mathbf{\Phi}),\text{rank}(\bar{\mathbf{G}}))+\text{rank}(\mathbf{H}_{{A,K_u}})\\
		&\leq \min(\min(\text{rank}(\mathbf{H}_{{R,K_u}}),M),\text{rank}(\bar{\mathbf{G}}))+\text{rank}(\mathbf{H}_{{A,K_u}})\\&=\min(\text{rank}(\mathbf{H}_{{R,K_u}}),\text{rank}(\bar{\mathbf{G}}))+\text{rank}(\mathbf{H}_{{A,K_u}}),
	\end{split}
\end{equation}
where
\begin{equation}
	\begin{split}
		\label{eq14-7}
		\mathbf{H}_{{R,K_u}}&=[\bar{\mathbf h}_{R,0},\bar{\mathbf h}_{R,1},\ldots, \bar{\mathbf h}_{R,K_u-1}]^T\in\mathbb{C}^{K_u\times M},\\
		\mathbf{H}_{{A,K_u}}&=[\bar{\mathbf h}_{A,0},\bar{\mathbf h}_{A,1},\ldots, \bar{\mathbf h}_{A,K_u-1}]^T\in\mathbb{C}^{K_u\times N}.
	\end{split}
\end{equation}

According to \eqref{eq14-2-2}, since both $\bar{\mathbf{G}}$ and $\mathbf{H}_{{R,k}}$ are non-zero matrices, $\text{rank}(\bar{\mathbf{G}})\geq 1$, $\text{rank}(\mathbf{H}_{{R,K_u}})\geq 1$. In
LOS channel, $\text{rank}(\mathbf{H}_{{R,K_u}})= 1$, we have the following inequality
\begin{equation}
	\begin{split}
		\label{eq14-9}
		\text{DoF}_1\leq1+K_u.
	\end{split}
\end{equation}

For the near-field, in the LoS channel, due to $\text{rank}(\bar{\mathbf{G}})= 1$, the highest degree of freedom is the same as in the far-field. Therefore, the deployment of IRS can reduce the maximum degree of freedom at Eve to 2, and for multi-user scenarios, the degree of freedom at users is greater than the Eve.

\subsection{BER Analysis}

Using quadrature phase shift keying (QPSK) modulation symbols, the receiver decodes the signal according to the adjusted constellation pattern. The bit error rate (BER) of the receiver can be calculated by \cite{Ding2014c}
\begin{equation}
	\begin{split}
		\label{eq:BER}
		\text{BER}=\frac{1}{4}\sum_{i=1}^{4}Q(\frac{L_i^2\sin(\alpha_i)}{N_0/2}),
	\end{split}
\end{equation}
where $L_i$ is the amplitude of the received signal, $\alpha_i$ is the minimum angle between the received signal vector and the IQ axis in the constellation diagram, $N_0/2$ is the power spectral density of AWGN, and $Q(\ )$ is the complementary cumulative distribution function (CCDF).

\section{Proposed UAV position and digital weight vector optimization method}\label{3}
In this section, in order to solve problem \eqref{eq20}, we first present the position optimization and digital weight vector optimization methods for UAV.
\subsection{Position optimization of UAV-DM system}

The appropriate power budget $P_\mathrm{max}$ is an important factor for the solution of $P2$ and $P3$~\cite{Zhou2019,Liu2020}. Given the UAV position and IRS reflection coefficient matrix $\boldsymbol{\Phi}$, the minimum transmit power $P_\mathrm{min}\leq P_\mathrm{max}$ that meets the requirements of DM design and minimum receiving sensitivity $\hat{t}_{b,k}=r_\mathrm{min}+\Delta_k/$tan$\varpi$ is given by
\begin{equation}
	\begin{split}
		\label{eq24}
	S1:&\operatorname*{min}_{\mathbf{w}_{b}}\parallel\mathbf{w}_{b}\parallel^{2} \\
	\mathrm{s.t.}\ &C1^{\prime\prime}:\mathbf{h}_{k}^{LoS}\mathbf{w}_{b}=(r_{\mathrm{min}}+\Delta_{k}/\tan\varpi)e^{j\vartheta_{b,k}},\forall k, \\
	&C2:\mathbf{h}_{e}^{LoS}\mathbf{w}_{b}=t_{b,e}e^{j\vartheta_{b,e}},\forall e. 
	\end{split}
\end{equation}
Note that the subproblem $S1$ is equivalent to the power required to meet the minimum requirements of the system, which is a convex optimization problem, and can be solved by the Lagrange multiplier method, given by
\begin{equation}
	\begin{split}	\label{eq25}
\mathcal{L}_{1}=\parallel\mathbf{w}_{b}\parallel^{2}-{\bm{\mu}}(\tilde{\mathbf{H}}^T\mathbf{w}_{b}-\tilde{\mathbf{C}}).
	\end{split}
\end{equation}
where $\bm{\mu}\in\mathbb{C}^{1\times(K_{u}+1)}$ is~the~Lagrange~multiplier,
\begin{equation}
	\begin{split}	\label{eq26}
\begin{aligned}
	\tilde{\mathbf{H}} &=[(\mathbf{h}_{0}^{LoS})^T,...,(\mathbf{h}_{k_{u}-1}^{LoS})^T,(\mathbf{h}_{e}^{LoS})^T] \\
	&=[\upsilon^{1}\zeta_{0}^{1}\overline{\mathbf{G}}^{T}\mathbf{\Phi}\overline{\mathbf{h}}_{R,0}+\zeta_{0}^{3}\overline{\mathbf{h}}_{A,0},..., \\
	&\quad\ \upsilon^{1}\zeta_{K_{u}-1}^{1}\bar{\mathbf{G}}^{T}\mathbf{\Phi}\bar{\mathbf{h}}_{R,K_{u}-1}+\zeta_{K_{u}-1}^{3}\mathbf{\bar{h}}_{A,K_{u}-1}, \\
	&\quad\ \upsilon^{1}\zeta_{e}^{1}\overline{\mathbf{G}}^{T}\mathbf{\Phi}\overline{\mathbf{h}}_{R,e}+\zeta_{e}^{3}\overline{\mathbf{h}}_{A,e} ]\in\mathbb{C}^{N\times(K_{u}+1)}, \\
	\tilde{\mathbf{C}} &=[(r_{\mathrm{min}}+\Delta_{0}/\tan\varpi)e^{j\vartheta_{b,0}},(r_{\mathrm{min}}+\Delta_{1}/\tan\varpi)e^{j\vartheta_{b,1}} \\
	&...,(r_{\mathrm{min}}+\Delta_{K_{u}-1}/\tan\varpi)e^{j\vartheta_{b,K_{u}-1}},t_{e}e^{j\vartheta_{b,e}}]^T\in\mathbb{C}^{(K_{u}+1)\times1}.
\end{aligned}
	\end{split}
\end{equation}
By setting $\partial\mathcal{L}_1/\partial\mathbf{w}_b^T=0$, we can obtain:
\begin{equation}
	\begin{split}	\label{eq27}
\mathbf{w}_b=\frac{1}{2}\bm{\mu}\tilde{\mathbf{H}}^T.
	\end{split}
\end{equation}
Substituting \eqref{eq27} into the constraint condition $C1^{\prime\prime}$ and $C2^{\prime}$ yields $\mathbf{w}_b$ as
\begin{equation}
	\begin{split}	\label{eq28}
		\mathbf{w}_b=\tilde{\mathbf{H}}(\tilde{\mathbf{H}}^T\tilde{\mathbf{H}})^{-T}\tilde{\mathbf{C}},
	\end{split}
\end{equation}
Then, according to \eqref{eq28}, the minimum power $P_\mathrm{min}$ required by the system can be obtained, and $\hat{t}_{b,k}$ can be increased in proportion to $\sqrt P_{\max}/\|\mathbf{w}_b\|.$

Since the positions of IRS and receiver $g$ are fixed, $\zeta_g^1\boldsymbol{\Phi h}_{R,g}^T$ is regarded as unchanged, and the position of UAV affects $\upsilon^1\overline{\mathbf{G}}$ and $\zeta_g^3\overline{\mathbf{h}}_{A,g}.$ According to \eqref{eq28}, we have:
\begin{equation}
	\begin{split}	\label{eq29}
\begin{aligned}
	\bar{\mathbf{h}}_{R,g}^T\mathbf{\Phi}\bar{\mathbf{G}}\mathbf{w}_{b}& =\bar{b}_{A,g}, \\
	\overline{\mathbf{h}}_{A,g}^T\mathbf{w}_{b}& =b_{A,g}, 
\end{aligned}
	\end{split}
\end{equation}
where $\bar{b}_{A,g}$ and $b_{A,g}$ are the beam gain of UAV-IRS-user (UAV-IRS-Eve) and UAV-User (UAV-Eve), respectively. Then, the UAV position optimization problem is formulated as
\begin{equation}
	\begin{split}	\label{eq30-sup}
		S2:\ &\max_\mathbf{u}\sum_{k=0}^{K_u-1}|\upsilon^1\zeta_k^1\bar{b}_{A,k}+\zeta_k^3b_{A,k}|\\
		s.t.\ &C6.
	\end{split}
\end{equation}
In accordance with the triangle inequality, we have
\begin{equation}
	\begin{split}	\label{eq30}
\sum_{k=0}^{K_u-1}|\upsilon^1\zeta_k^1\bar{b}_{A,k}+\zeta_k^3b_{A,k}|\leq\sum_{k=0}^{K_u-1}(\upsilon^1\zeta_k^1|\bar{b}_{A,k}|+\zeta_k^3|b_{A,k}|).\\
	\end{split}
\end{equation}
Then, the UAV position optimization problem can be reformulated as
\begin{equation}
	\begin{split}	\label{eq30-1}
		S3:\ &\max_\mathbf{u}\sum_{k=0}^{K_u-1}(\upsilon^1\zeta_k^1|\bar{b}_{A,k}|+\zeta_k^3|b_{A,k}|)\\
		s.t.\ &C6.
	\end{split}
\end{equation}
Furthermore, the objective function in \eqref{eq30-1} is equivalent to
\begin{equation}
	\begin{split}	\label{eq31}
\min_{\mathbf{u}}\ J_1=\frac1{\sum_{k=0}^{K_u-1}(\frac{\sqrt{\rho\varepsilon_{\tiny{A,R}}}}{d_{A,R}}\zeta_k^1|\bar{b}_{A,k}|+\frac{\sqrt{\rho\varepsilon_{\tiny{A,k}}}}{d_{A,k}}|b_{A,k}|)}.
	\end{split}
\end{equation}
In accordance with the arithmetic-geometric means inequality, i.e.,
\begin{equation}
	\begin{split}	\label{eq32}
\begin{aligned}\frac1{\sum_{i=1}^na_i}\leq\frac1n\frac1{\left(\prod_{i=1}^na_i\right)^{1/n}}&=\frac1n(\prod_{i=1}^n\frac1{a_i})^{\frac1n}\\&\leq\frac1{n^2}\sum_{i=1}^n\frac1{a_i}\leq\sum_{i=1}^n\frac1{a_i}.\end{aligned}
	\end{split}
\end{equation}
With \eqref{eq32}, \eqref{eq31} is upper bounded by
\begin{equation}
	\begin{split}	\label{eq33}
\begin{aligned}
	J_{1}& \leq\sum_{k=0}^{{K_{u}-1}}(\frac{d_{A,R}}{\sqrt{\rho\varepsilon_{A,R}}\zeta_{k}^{1}|\bar{b}_{A,k}|}+\frac{d_{A,k}}{\sqrt{\rho\varepsilon_{A,k}}|b_{A,k}|}) \\
	&=\sum_{k=0}^{K_u-1}(d_{A,R}T_k^1+d_{A,k}T_k^2),
\end{aligned}
	\end{split}
\end{equation}
where
\begin{equation}
	\begin{split}	\label{eq34}
\begin{aligned}&T_{k,1}=\frac{1}{\sqrt{\rho\varepsilon_{A,R}}\zeta_{k}^{1}|\bar{b}_{A,k}|},\\&T_{k,2}=\frac{1}{\sqrt{\rho\varepsilon_{A,k}}|b_{A,k}|}.\end{aligned}
	\end{split}
\end{equation}
Therefore, the approximate solution obtained by minimizing this upper bound can be given by
\begin{equation}
	\begin{split}	\label{eq35}
\min_\mathbf{u}J_2=\sum_{k=0}^{K_u-1}(d_{A,R}T_{k,1}+d_{A,k}T_{k,2}).
	\end{split}
\end{equation}
By converting spherical coordinates to Cartesian coordinates, with $\theta_{A,R}, \varphi_{A,R}, \theta_{R,k}$ and $\varphi_{R,k}$, the Cartesian coordinates of UAV and users can be represented as $(x_{A},y_{A},H_{A}-H_{R})$ and $(x_{k},y_{k},-H_{R})$, respectively. For \eqref{eq35}, the derivative of $J_2$ with respect to $x_A$ and $y_A$ respectively yields
\begin{equation}
	\begin{split}	\label{eq36}
\partial J_2 / \partial x_A=\sum_{k=1}^K((x_A-x_R)d_{A,R}^{-1}T_{k,1}+(x_A-x_k)d_{A,k}^{-1}T_{k,2}),\\\partial J_2 / \partial y_A=\sum_{k=1}^K((y_A-y_R)d_{A,R}^{-1}T_{k,1}+(y_A-y_k)d_{A,k}^{-1}T_{k,2}).
	\end{split}
\end{equation} 
Correspondingly, the second-order derivative of $J_2$ is expressed as 
\begin{equation}
	\begin{split}	\label{eq37}
\begin{aligned}
	\frac{\partial^2J_2}{\partial x_A^2}& =\sum_{k=1}^{K}d_{A,R}^{-1}T_{k,1}+d_{A,k}^{-1}T_{k,2} \\
	&-\sum_{k=1}^K((x_A-x_R)^2d_{A,R}^{-3}T_{k,1}+(x_A-x_k)^2d_{A,k}^{-3}T_{k,2}), \\
	\frac{\partial^2J_2}{\partial y_A^2}& =\sum_{k=1}^Kd_{A,R}^{-1}T_{k,1}+d_{A,k}^{-1}T_{k,2} \\
	&-\sum_{k=1}^K((y_A-y_R)^2d_{A,R}^{-3}T_{k,1}+(y_A-y_k)^2d_{A,k}^{-3}T_{k,2}), \\
	\frac{\partial^2J_2}{\partial x_Ay_A}& =\frac{\partial^2J_2}{\partial y_Ax_A}=-\sum_{i=1}^K((x_A-x_R)(y_A-y_R)d_{A,k}^{-3}T_{k,1} \\
	&-\sum_{i=1}^K((x_A-x_k)(y_A-y_k)d_{A,k}^{-3}T_{k,2}).
\end{aligned}
	\end{split}
\end{equation}  
Then, the Hessian matrix of $J_2$ is given as 
\begin{equation}
	\begin{split}	\label{eq38}
\mathbf{H}_{J_2}=\begin{bmatrix}\frac{\partial^2J_2}{\partial x_A^2}&\frac{\partial^2J_2}{\partial x_Ay_A}\\\frac{\partial^2J_2}{\partial y_Ax_A}&\frac{\partial^2J_2}{\partial y_A^2}\end{bmatrix}.
	\end{split}
\end{equation}   
The determinant of $\mathbf{H}_{J_2}$ can be proven to be positive using Cauchy-Schwarz inequality. Therefore, \eqref{eq35} is convex, and by setting $\partial J_2/\partial x_A=0$ and $\partial J_2/\partial y_A=0$, we can obtain
\begin{equation}
	\begin{split}	\label{eq39}
x_{A}\sum_{k=1}^{K}(d_{A,R}^{-1}T_{k,1}+d_{A,k}^{-1}T_{k,2})=\sum_{k=1}^{K}(x_Rd_{A,R}^{-1}T_{k,1}+x_{k}d_{A,k}^{-1}T_{k,2}),\\
y_{A}\sum_{k=1}^{K}(d_{A,R}^{-1}T_{k, 1}+d_{A,k}^{-1}T_{k,2})=\sum_{k=1}^{K}(y_Rd_{A,R}^{-1}T_{k,1}+y_{k}d_{A,k}^{-1}T_{k,2}).
	\end{split}
\end{equation}  
Then, the fixed point iteration method can be used to solve \eqref{eq39}. Furthermore, considering the impact of changes in UAV position on $\mathbf{w}_b$, recalculate \eqref{eq28} and \eqref{eq29}, and optimize UAV position $\mathbf{u}$ again. Note that if $||\mathbf{w}_b||$ increases, it means that higher transmit power is required to meet the minimum system requirements, and the iteration ends.

\subsection{Optimization of digital weight vector}

Given the position of UAV and set the transmit power $P_\mathrm{max}=P_\mathrm{min}$, the subproblem with respect to the digital weight vector $\mathbf{w}_{b}$ is formulated as
\begin{equation}
	\begin{split}	\label{eq40}
		S4:&J_3 =\operatorname*{max}_{\mathbf{w}_{b}}\sum_{k=0}^{K_u-1}\hat{t}_k\\
		\mathrm{s.t.}\ &C1^{\prime}, C2^{\prime}, C3^{\prime},\\
		&C7:\parallel\mathbf{w}_b\parallel^2=P_{\min}.
	\end{split}
\end{equation} 
 
To solve this problem, the penalty function method is used to represent $S4$ as
\begin{equation}
	\begin{split}	\label{eq41}
	S5:&J_4 =\operatorname*{max}_{\mathbf{w}_{b},\{\hat{t}_{b,k}\}}\sum_{k=0}^{K_u-1}\hat{t}_{b,k}-\xi(\sum_{k=0}^{K_u-1}||\hat{t}_{b,k}e^{j\vartheta_{b,{k}}}-\mathbf{h}_{k}^{LoS}\mathbf{w}_{b}||^2\\
	&\quad+||{t}_{b,e}e^{j\vartheta_{b,{e}}}-\mathbf{h}_{e}^{LoS}\mathbf{w}_{b}||^2+(\parallel\mathbf{w}_b\parallel^2-P_{\min})^2)\\
\mathrm{s.t.}\ &C3^{\prime},
	\end{split}
\end{equation}  
where $\xi$ is the penalty variable. Given the $\mathbf{w}_{b}$, \eqref{eq41} is simplified as the one with respect to 
$\hat{t}_{b,k}$ only. Then, calculate ${\partial J_4} /{\partial \hat{t}_{b,k}}$ as
\begin{equation}
	\begin{split}	\label{eq42}
\frac{\partial J_4} {\partial \hat{t}_{b,k}}=1-2\xi\hat{t}_{b,k}+\xi ((\mathbf{h}_{k}^{LoS}\mathbf{w}_{b})^He^{j\vartheta_{b,{k}}}+\mathbf{h}_{k}^{LoS}\mathbf{w}_{b}e^{-j\vartheta_{b,{k}}}). \\
	\end{split}
\end{equation} 
By setting ${\partial J_4} /{\partial \hat{t}_{b,k}}=0$, we can obtain
\begin{equation}
	\begin{split}	\label{eq43}
\hat{t}_{b,k}=\frac{1+\xi ((\mathbf{h}_{k}^{LoS}\mathbf{w}_{b})^He^{j\vartheta_{b,{k}}}+\mathbf{h}_{k}^{LoS}\mathbf{w}_{b}e^{-j\vartheta_{b,{k}}})}{2\xi}.
	\end{split}
\end{equation} 
With $C3^{\prime}$, we have
\begin{equation}
	\begin{split}	\label{eq44}
	{\xi}\leq \frac{1}{2r_{min,k}-((\mathbf{h}_{k}^{LoS}\mathbf{w}_{b})^He^{j\vartheta_{b,{k}}}+\mathbf{h}_{k}^{LoS}\mathbf{w}_{b}e^{-j\vartheta_{b,{k}}})}.
	\end{split}
\end{equation} 

With the obtained $\hat{t}_{b,k}$, the subproblem with respect to $\mathbf{w}_{b}$ can be formulated as
\begin{equation}
	\begin{split}	\label{eq45}
		S6:&J_5 =\operatorname*{max}_{\mathbf{w}_{b}}\ -(\sum_{k=0}^{K_u-1}||\hat{t}_{b,k}e^{j\vartheta_{b,{k}}}-\mathbf{h}_{k}^{LoS}\mathbf{w}_{b}||^2\\
		&\quad+||{t}_{b,e}e^{j\vartheta_{b,{e}}}-\mathbf{h}_{e}^{LoS}\mathbf{w}_{b}||^2+(\parallel\mathbf{w}_b^T\parallel^2-P_{\min})^2),
	\end{split}
\end{equation}  
which can be calculated through \eqref{eq28}. According to \eqref{eq43} and \eqref{eq28}, iteratively optimize $\hat{t}_k$ and $\mathbf{w}_{b}$ until the constraint $C4$ is not satisfied and the iteration ends.

\section{Optimization of reflection coefficient matrix}\label{4}

In this section, we propose three methods to optimize the IRS phase shift $\mathbf{\Phi}$ matrix based on the given UAV position and digital weight vector $\mathbf{w}_b$, in order to achieve the maximum transmission rate.
\subsection{Proposed vector trajectory (VT) method}

\begin{figure}
\centering
\includegraphics[width=0.30\textwidth, trim = 2 2 2 2,clip]{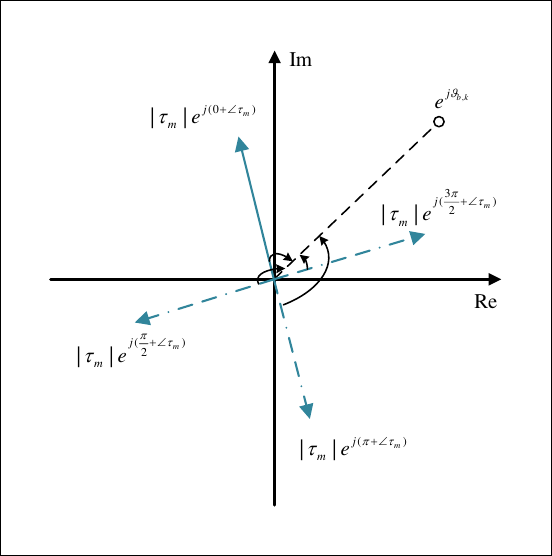}\\
\caption{An example of phase selection according to the VT method.}\label{fig:VT}
\end{figure}

With $\mathbf{u}$ and $\mathbf{w}_b$, the subproblem with respect to $\mathbf{\Phi}$ is formulated as
\begin{equation}
	\begin{split}	\label{eq46}
		S7:&J_6 =\operatorname*{max}_{\mathbf{\Phi}}\sum_{k=0}^{K_u-1}\hat{t}_{b,k}\\
		\mathrm{s.t.}\ &C1^{\prime}, C2^{\prime}, C3^{\prime}, C5^{\prime}.\\
	\end{split}
\end{equation} 
For \eqref{eq46}, the sum of user transmission rates is improved by optimizing $\mathbf{\Phi}$. To solve this subproblem, a vector trajectory (VT) method is proposed from the perspective of adding multiple complex numbers.

In \eqref{eq29}, both $\bar{b}_{A,g}$ and ${b}_{A,g}$ are complex values, which are obtained by summing a series of complex terms. With $\gamma_m=1$, $\bar{b}_{A,g}$ can be represented as
\begin{equation}\label{eq51}
	\begin{split}
		\bar{b}_{A,g}&=\sum_{m=0}^{M-1}\bar{b}_{A,R,m}\bar{{h}}_{R,g,m}e^{j\phi_m}=\sum_{m=0}^{M-1}|\tau_m|e^{j(\phi_m+\angle{\tau_m})},
	\end{split}
\end{equation}
where
\begin{equation}\label{eq52}
	\begin{split}
		\bar{\mathbf{h}}_{R,g}&=[\bar{{h}}_{R,g,0},\bar{{h}}_{R,g,1},\ldots,\bar{{h}}_{R,g,M-1}]^T,\\
		\bar{\textbf{b}}_{A,R}&=\bar{\mathbf{G}}\mathbf{w}_{b}=[\bar{b}_{A,R,0},\bar{b}_{A,R,1},\ldots,\bar{b}_{A,R,M-1}]^T,\\
		\tau_m&=\bar{b}_{A,R,m}\bar{{h}}_{R,g,m}.
	\end{split}
\end{equation}

Given $\textbf{u}$, $\mathbf{w}_{b}$, and $\gamma_m=1$, ${b}_{A,g}$ and $\tau_m$ are fixed. At the position of user $k$, the projection of the unit symbol $e^{j\vartheta_{b,{k}}}$ by $\bar{b}_{A,g}$ can be represented as
\begin{equation}\label{eq53}
	\begin{split}
		\text{Proj}_{e^{j\vartheta_{b,{k}}}}\bar{b}_{A,g}&=\sum_{m=0}^{M-1}|\tau_m|\cos{(\phi_m+\angle{\tau_m}-\vartheta_{b,{k}})}\\&=\sum_{m=0}^{M-1}|\tau_m|\cos\psi_m,
	\end{split}
\end{equation}
where $\psi_m={\phi_m+\angle{\tau_m}-\vartheta_{b,{k}}}$, $-\pi/2\leq\psi_m\leq\pi/2$. According to \eqref{eq53}, the smaller $|\psi_m|$, the greater the beam gain $\bar{b}_{A,g}$, which can achieve a higher user transmission rate. Accordingly, for single user scenarios, the phase shift $\phi_m$ of the $m$-th unit of the IRS can be optimized by
\begin{equation}\label{eq54}
	\begin{split}
		\phi_m^*=\text{arg}\min_{\phi_m\in\mathbb{F}}|\phi_m+\angle{\tau_m}-\vartheta_{b,{k}}|=\text{arg}\min_{\phi_m\in\mathbb{F}}|\psi_m|.
	\end{split}
\end{equation}
Taking $\tilde{B}=2$ as an example, as shown in Fig. \ref{fig:VT}, the $m$-th vector approaches the imaginary axis in the second quadrant, with an expected symbol of $e^{j\vartheta_{b,{k}}}$. The angle between the rotation of the $m$-th vector by $3\pi/2$ and the expected symbol is the smallest. Therefore, the optimal phase shift of this electromagnetic unit is $3\pi/2$.

Correspondingly, the worst gain $\hat{b}_{A,g}$ related to phase shift accuracy $\tilde{B}$ is
\begin{equation}\label{eq55}
	\begin{split}
		\hat{b}_{A,g}=\sum_{m=0}^{M-1}|\tau_m|\cos\frac{2\pi}{2^{\tilde{B}+1}}.
	\end{split}
\end{equation}
With $\tilde{B}\geq1$, $0\leq|\psi_m|\leq\pi/2$, and then $\hat{b}_{A,g}\geq0$. High precision phase shifters are beneficial for achieving higher gains.

For multi-user scenarios, since multiple users share the same IRS, the optimal phase shift set $\bm{\phi}_k=[\phi_{0,k},\phi_{1,k},\ldots,\phi_{M-1,k}]^T$ for each legitimate user is calculated according to \eqref{eq54}. The phase shift values are prioritized from high to low. For example, for the $k$-th user, based on the gain obtained, for $m$-th unit, if the value of selecting the first element in $\mathbb{F}$ is greater than the value of when it is the zeroth element, then for user $k$, selecting the zeroth element in $\mathbb{F}$ for the $m$-th unit has a higher priority than choosing the first element. Then, according to the priority of different elements corresponding to users $k$ and IRS units, select the phase with the same priority as the corresponding element as the optimal phase for that unit. For units with different priorities, flexible design can be used to disturb the eavesdropper. According to \eqref{eq55}, the gain of the IRS is greater than or equal to zero, which means there is no gain or forward gain for the user. 

Specifically, introducing matrix $\bm{\Phi}_k$ corresponding to the $k$-th user, i.e.,
\begin{equation}\label{eq56}
	\begin{split}
		\bm{\Phi}_k=
		\begin{bmatrix}
			\phi_{k,0,0}&\phi_{k,0,1}&\cdots&\phi_{k,0,M-1}\\
			\phi_{k,1,0}&\phi_{k,1,1}&\cdots&\phi_{k,1,M-1}\\
			\vdots&&\ddots&\vdots\\
			\phi_{k,2^{\tilde{B}-1},0}&\phi_{k,2^{\tilde{B}-1},1}&\cdots&\phi_{k,2^{\tilde{B}-1},M-1}
		\end{bmatrix},
	\end{split}
\end{equation}
where the columns represent IRS units, the rows represent priorities, and the phase priorities decrease with increasing row numbers. $\phi_{k,\bar{p},m}=\text{arg}\min_{\phi_{k,\bar{p},m}\in\mathbb{F}}|\phi_{k,\bar{p},m}+\angle{\tau_m}-\vartheta_{b,{k}}|$ represents the phase corresponding to user $k$ at the $m$-th IRS unit with priority $\bar{p}=0,1,\ldots,2^{\tilde{B}}-1$. Then, for the $m$-th IRS unit, the same phase set ${\mathcal B}_m$ corresponding to the same priority for all users can be represented as
\begin{equation}\label{eq57}
	{\mathcal B}_m=\{\phi_{k,\tilde{p},m}\}_{\tilde{p}\in\mathcal P_m}, 
\end{equation}
where ${\mathcal P}_m$ is the priority set corresponding to the same phase at the $m$-th IRS unit at the $m$-th IRS unit. Then, the optimal phase in multi-user scenarios can be represented as
\begin{equation}\label{eq58}
	\phi_m^*=\text{arg}\min_{\phi_{k,\tilde{p},m}\in\mathbb{B}_m}|\phi_{k,\tilde{p},m}+\angle{\tau_m}-\vartheta_{b,{k}}|.
\end{equation}

Correspondingly, the specific steps of the optimization method are summarized in Algorithm \ref{alg:alg2}.  $\textbf{w}_b^{(l)}$, $x_A^{(l)}$, and $y_A^{(l)}$ represent the values obtained after the $l$-th iteration.
\begin{algorithm}
	\caption{Joint design of $\mathbf{w}_b$, $\mathbf{u}$ and $\mathbf{\Phi}$ with VT method.}\label{alg:alg2}
	\begin{algorithmic}[1]
		\STATE \textbf{Initialize}: Iteration index $l=0$, $\textbf{w}_b^{(0)}$, $\mathbf{\Phi}^{(0)}$
		\STATE\textbf{repeat}
		\STATE\hspace{0.5cm}$l=l+1$;
		\STATE\hspace{0.5cm}Use fixed point iteration method to solve \eqref{eq39};
		\STATE\hspace{0.5cm}Update $\hat{t}_{k}$ according to \eqref{eq43};
		\STATE\hspace{0.5cm}Update $\textbf{w}_b^{(l)}$ according to \eqref{eq28};
		\STATE\textbf{until}$||\textbf{w}_b^{(l+1)}||^2-P_\mathrm{max}>0$
		\STATE Calculate $\phi_m^*$ according to \eqref{eq58}.
		\STATE\textbf{Output}: Suboptimal solution \{${\phi}_{m}^*$, $\textbf{w}_b^{(l)}$, $x_A^{(l)}$, $y_A^{(l)}$\}
	\end{algorithmic}
\end{algorithm}

The proposed VT method optimizes $\bm{\Phi}_{m}$ based on the given $\textbf{w}_b$ and $\textbf{u}$. Using the VT method, a larger $\bar{b}_{A,R,m}$ is beneficial for achieving higher power gain. Correspondingly, the cross entropy VT (CE-VT) and block coordinate descent VT (BCD-VT) are proposed.

\subsection{Proposed CE-VT algorithm}

To solve \eqref{eq46}, we introduce the cross entropy (CE) method. For the $b$-th symbol, the minimum cost function value is unknown, i.e., the optimal sample of $\bm{\phi}$ is uncertain, where $\bm{\phi}=[\phi_{0}, \phi_{1}, \ldots, \phi_{M-1}]^T$ is the phase of IRS. Specifically, a probability matrix $\textbf{P}^{(i)} \in \mathbb{C}^{2^{\tilde{B}} \times M}$ is introduced to randomly generate $X$ samples of $\bm{\phi}$, and $X_{e}$ elite samples are selected from them to estimate a new probability matrix $\textbf{P}^{(i+1)}$ to generate better samples, where $i$ is the iteration index. The probability matrix for the $i$-th iteration can be expressed as
\begin{equation}
	\begin{split}
		\textbf{P}^{(i)} &= [\textbf{p}^{(i)}_{0}, \textbf{p}^{(i)}_{1}, \ldots, \textbf{p}^{(i)}_{M-1}],\\
		\mbox{with}\quad \textbf{p}^{(i)}_{m} &= [p^{(i)}_{m,0}, p^{(i)}_{m,1}, \ldots, p^{(i)}_{m,\tilde{B}-1}]^{T},\\
		for\quad\quad\quad i &= 0, 1, \ldots, \bar{I}-1,
	\end{split}
\end{equation}
where $p^{(i)}_{m,\tilde{b}}$ $(\tilde{b}=0, 1, \ldots, 2^{\tilde{B}}-1)$ is the probability of the $\tilde{b}$-th element in $\mathbb{F}$ being taken, $\bar{I}$ is the number of iterations.

Assume that all samples and all elements of the same sample are independent of each other, and for the $m$-th element, all elements in $\textbf{p}_{i,m}$ have values between 0 and 1 and the sum is 1, i.e. $\sum_{\tilde{b}=0}^{2^{\tilde{B}-1}}p^{(i)}_{m,\tilde{b}}=1$ and $0 \leqslant p^{(i)}_{m,\tilde{b}} \leqslant 1$. Then, the probability distribution function $G(\bm{\phi}; \textbf{P}^{(i)})$ can be expressed as~\cite{Chen2019b}
\begin{equation}\label{eq:probability distribution}
	\begin{split}
		&G(\bm{\phi}; \textbf{P}^{(i)}) = \prod_{m=0}^{M-1}\Bigg(\prod_{\tilde{b}=0}^{2^{\tilde{B}}-1}(p^{i}_{m,\tilde{b}})^{H(\phi_{m},F(\tilde{b}))}\Bigg),\\
		&\mbox{with}\quad H(\phi_{m},F(\tilde{b}))=\begin {cases}
		1  &\text {$\phi_{m}=F(\tilde{b})$}  \\
		0 &\text {$\phi_{m}\neq F(\tilde{b})$},
	\end{cases}
\end{split}
\end{equation}
where $F(\tilde{b})$ is the $\tilde{b}$-th entry of $\mathbb{F}$. $G(\bm{\phi}; \textbf{P}^{(i)})$ means that the probability that sample $\bm{\phi}$ is taken when the probability matrix is $\textbf{P}^{(i)}$. In order to obtain better samples, $X$ samples are sorted from large to small based on the objective function value, and the top $X_e$ samples corresponding to the values are selected as elite samples. Then, the probability matrix $\textbf{P}^{(i+1)}$ of the next iteration can be updated by
\begin{equation}\label{eq50}
\begin{split}
	&\underset{\textbf{P}^{(i+1)}}{\text{max}}
	X^{-1}\sum_{\bar{x}=0}^{X_{e}-1}\ln G(\bm{\phi}_{\bar{x}}; \textbf{P}^{(i+1)})\\
	\text{s.t.}\ &\sum_{\tilde{b}=0}^{2^{\tilde{B}}-1}p^{(i)}_{m,\tilde{b}}=1, \;\;0 \leqslant p^{(i)}_{m,\tilde{b}} \leqslant 1
\end{split}
\end{equation}
where $\bm{\phi}_{\bar{x}}$ represents the $\bar{x}$-th sample. \eqref{eq50} is a convex optimization problem that can be solved using the CVX toolbox \cite{research12a}. Then, iteratively optimize $\textbf{w}_b$, $\mathbf{u}$, and $\bm{\Phi}$, where $\bm{\Phi}=diag(\bm{\phi}_{\text{max}})$, $\bm{\phi}_{\text{max}}$ is the sample corresponding to the maximum objective function value. 

According to the CE method, higher $\bar{b}_{A,R,m}$ can be obtained, and \eqref{eq58} can be used to increase the transmission rate. Correspondingly, the specific steps of the optimization method are summarized in Algorithm \ref{alg:alg1}. $\textbf{w}_b^{(i)}$, $x_A^{(i)}$, and $y_A^{(i)}$ represent the values obtained after the $i$-th iteration.
\begin{algorithm}
	\caption{Joint design algorithm of $\mathbf{w}_b$, $\mathbf{u}$ and $\mathbf{\Phi}$ with CE-VT method.}\label{alg:alg1}
	\begin{algorithmic}[1]
		\STATE \textbf{Initialize}: Iteration index $i=0$, $\textbf{w}_b^{(0)}$, $\textbf{P}^{(0)}$
		\STATE\textbf{repeat}
		\STATE\hspace{0.5cm}$i=i+1$;
		\STATE \hspace{0.5cm}Obtain  $X$ samples
		$\{\bm{\phi}_{\bar{x}}\}^{X-1}_{\bar{x}=0}$ through $\textbf{P}^{(i)}$;
		\STATE\hspace{0.5cm}\textbf{repeat}
		\STATE\hspace{0.5cm}\hspace{0.5cm}$l=l+1$;
		\STATE\hspace{0.5cm}\hspace{0.5cm}Use fixed point iteration method to solve \eqref{eq39};
		\STATE\hspace{0.5cm}\hspace{0.5cm}Update $\hat{t}_{k}$ according to \eqref{eq43};
		\STATE\hspace{0.5cm}\hspace{0.5cm}Update $\textbf{w}_b^{(l)}$ according to \eqref{eq28};
		\STATE \hspace{0.5cm}\textbf{until} $||\textbf{w}_b^{(l+1)}||^2-P_\mathrm{max}>0$
		\STATE \hspace{0.5cm}Update $\textbf{P}^{(i)}$ according to \eqref{eq50};
		\STATE\textbf{until} $\textbf{P}^{(i-2)}=\textbf{P}^{(i-1)}=\textbf{P}^{(i)}$
		\STATE Update $\bm{\phi}_{\text{max}}$ according to \eqref{eq58};
		\STATE\textbf{Output}: Suboptimal solution \{$\bm{\phi}_{\text{max}}$, $\textbf{w}_b^{(i)}$, $x_A^{(i)}$, $y_A^{(i)}$\}
	\end{algorithmic}
\end{algorithm}

The CE method, which learns the phase shift optimal distribution through multiple iterative sampling, has lower complexity than the exhaustive search method.

\subsection{Proposed BCD-VT algorithm}

By using the block coordinate descent (BCD) method, fixing $M-1$ reflection units and optimizing one unit, the discrete phase constraint problem can be solved.

Specifically, based on the initial phase shift matrix, $\mathbf{w}_b$ and $\mathbf{\Phi}$ can be obtained using $\eqref{eq28}$ and $\eqref{eq39}$. Firstly, optimize the zeroth unit and fix the other $M-1$ units. The phase of the zeroth unit is taken from the $\mathbb{F}$, i.e., the phase values in $\mathbb{F}$ are traversed to obtain $2^{\tilde{B}}$ objective function values. Update the phase of the zeroth element to the phase corresponding to the maximum objective function value. Then, optimize the $m$-th element and fix the $M-1$ elements until the last element is optimized.

\begin{algorithm}
	\caption{Joint design algorithm of $\mathbf{w}_b$, $\mathbf{u}$ and $\mathbf{\Phi}$ with BCD-VT method.}\label{alg:alg3}
	\begin{algorithmic}[1]
		\STATE \textbf{Initialize}: Iteration index $l=0$, $\tilde{b}=0$, IRS unit index $m=0$, $\textbf{w}_b^{(0)}$, $\mathbf{\Phi}^{(0)}$
		\STATE\textbf{repeat}
		\STATE\hspace{0.5cm}\textbf{repeat}
		\STATE \hspace{0.5cm}\hspace{0.5cm}Set the phase of the $m$-th IRS unit to $F(\tilde{b})$;
		\STATE\hspace{0.5cm}\hspace{0.5cm}$\tilde{b}=\tilde{b}+1$;
		\STATE\hspace{0.5cm}\hspace{0.5cm}\textbf{repeat}
		\STATE\hspace{0.5cm}\hspace{0.5cm}\hspace{0.5cm}$l=l+1$;
		\STATE\hspace{0.5cm}\hspace{0.5cm}\hspace{0.5cm}Use fixed point iteration method to solve \eqref{eq39};
		\STATE\hspace{0.5cm}\hspace{0.5cm}\hspace{0.5cm}Update $\hat{t}_{k}$ according to \eqref{eq43};\\
		\STATE\hspace{0.5cm}\hspace{0.5cm}\hspace{0.5cm}Update $\textbf{w}_b^{(l)}$ according to \eqref{eq28};
		\STATE\hspace{0.5cm}\hspace{0.5cm}\textbf{until} $||\textbf{w}_b^{(l+1)}||^2-P_\mathrm{max}>0$
		\STATE\hspace{0.5cm}\textbf{until} $\tilde{b}=\tilde{B}$\\
		\STATE\hspace{0.5cm}{Select the phase corresponding to the maximum objective function value as the optimal phase for the $m$-th unit.}
		\STATE\hspace{0.5cm}$m=m+1$;
		\STATE\textbf{until} $m=M$
		\STATE Calculate $\phi_m^*$ according to \eqref{eq58}.
		\STATE\textbf{Output}: Suboptimal solution \{${\phi}_{m}^*$, $\textbf{w}_b^{(l)}$, $x_A^{(l)}$, $y_A^{(l)}$\}
	\end{algorithmic}
\end{algorithm}
According to the BCD method, higher $\bar{b}_{A,R,m}$ can be obtained, and \eqref{eq58} can be used to increase the transmission rate. Correspondingly, the specific steps of the optimization method are summarized in Algorithm \ref{alg:alg3}.
	
\section{Simulation results}\label{5}

In this section, we evaluated the transmission rate and security performance of a multi-user UAV-DM system through simulation. The UAV is equipped with $N=24$ antennas, and the number of electromagnetic units on the IRS is $M=576$, where $M_Y=M_Z=24$. The initial position of the UAV is random. The IRS is 1 meter above the ground and its position is $(0m, 0m, 0m)$. The three ground users are located at $(10m, 15m, -1m)$, $(20m, 10m, -1m)$, and $(15m, 20m, -1m)$, respectively. The Eve's position is $(10m, 20m, -1m)$. The carrier frequency is $50GHz$, and $\delta^2=-110dBm$. The array aperture is 0.39m, and the Ruili distance is 50.75m. In simulation, the computer we used is 13-th Gen Intel (R) Core (TM) i7-13700KF 3.40 GHz and 32GB RAM. The main parameters are shown in Table \ref{1}.

\begin{table}[H]
	\centering
	\caption{THE SIMULATION PARAMETERS}
	\resizebox{\linewidth}{!}{
	\begin{tabular}{l|c}
		\toprule[0.7pt]
		\textbf{Parameters} & \textbf{Values}  \\
		\midrule[0.7pt]
		UAV height& $H_u = 100m$\\
		Number of antennas& $N = 24$\\
		Number of users& $K_u = 3$\\
		minimum symbol power& $r_min = -80dBm$\\
		represent the LoS power ratio coefficients& $\varepsilon_{{A, R}}=\varepsilon_{{R, g}}=\varepsilon_{A,g}=0.9$\\
		Channel power gain per unit distance&  $\rho = 10^{-3}$\\
		Power of noise&  $\delta^2=-110dBm$\\
		Location range&  $\theta_{A,R}\in[\frac{5\pi}{9}\ \frac{5\pi}{6}]$, $\phi_{A,R}\in[0\  \frac{\pi}{2}]$\\
		\bottomrule[0.5pt]
	\end{tabular}}
	\label{1}
\end{table}
\begin{figure}[H]
	\centering
	\includegraphics[width=0.45\textwidth, trim = 50 260 20 280,clip]{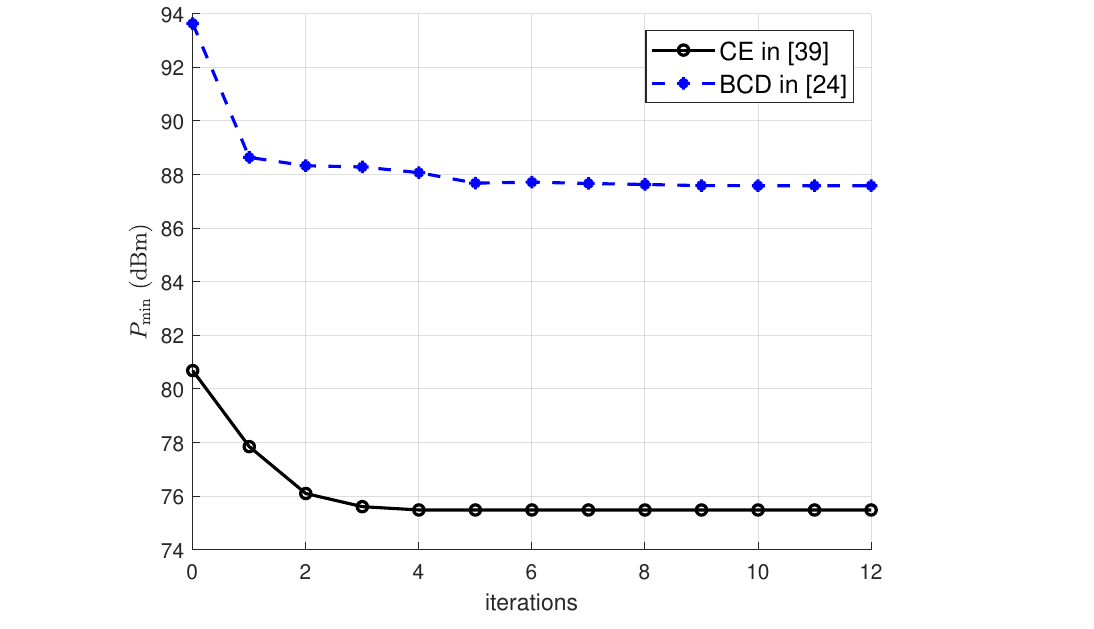}\\
	\caption{Minimum transmission power $P_{min}$ versus number of iterations.}\label{fig:S5}
\end{figure}
 
\begin{figure}[t]
	\centering
	\includegraphics[width=0.45\textwidth, trim = 50 260 20 280,clip]{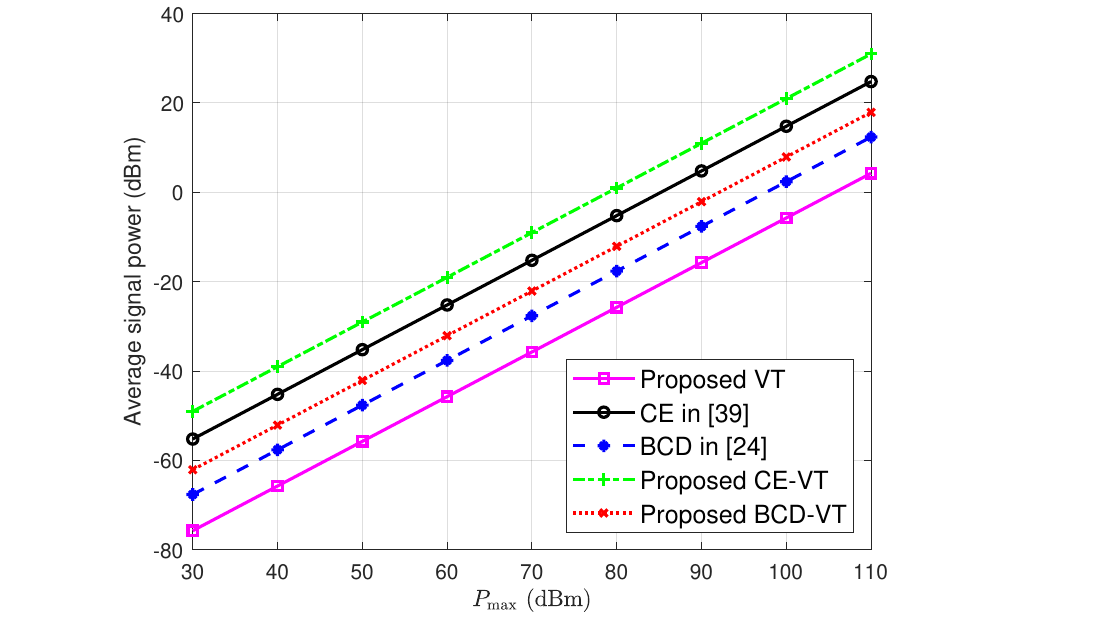}\\
	\caption{The average signal power of different methods.}\label{fig:S3}
\end{figure}

\begin{figure}[t]
	\centering
	\includegraphics[width=0.45\textwidth, trim = 50 260 20 280,clip]{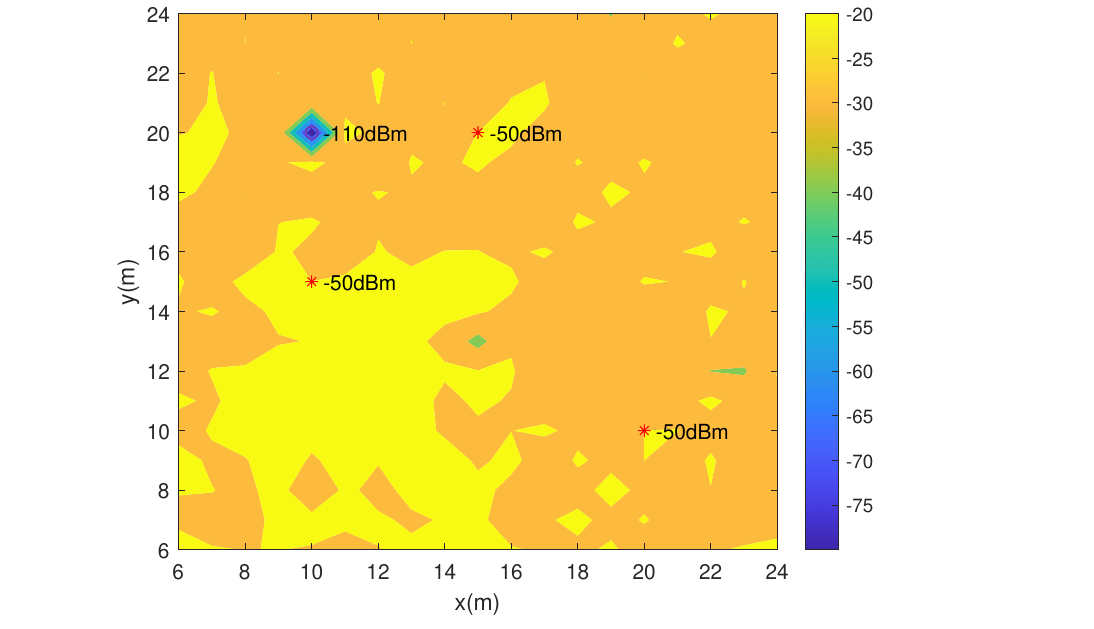 }\\
	\caption{Beam response. The asterisk represents the user, and the circle is the eavesdropper.}\label{fig:S1}
\end{figure}

The convergence curves under the CE method and BCD method are shown in Fig. \ref{fig:S5}. In Fig. \ref{fig:S3}, we plot the average signal power under different algorithms, comparing the vector trajectory (VT) method, cross entropy (CE) method, block coordinate descent (BCD) method, CE-VT algorithm, and BCD-VT algorithm. From Fig. \ref{fig:S3}, it can be observed that as $P_{max}$ increases, the CE-VT algorithm proposed can achieve a signal power improvement of about $6dB$ higher than the CE method, and the BCD-VT algorithm can achieve a signal power improvement of about $5.5dB$ higher than the BCD method, and has a higher performance improvement compared to the VT method. Fig. \ref{fig:S1} shows the beam responses obtained using the CE-VT algorithm, which represents the power distribution of the received signal within the region. With a maximum transmission power of $P_{max}=73.6dBm$, the power of the three users (red asterisk) is $-50dBm$, and the eavesdropping (red circle) position is $-110dBm$. The proposed method can achieve high power at the user's location and low power at the eavesdropper's location. 


\begin{figure}
	\centering
	\includegraphics[width=0.45\textwidth, trim = 50 260 20 280,clip]{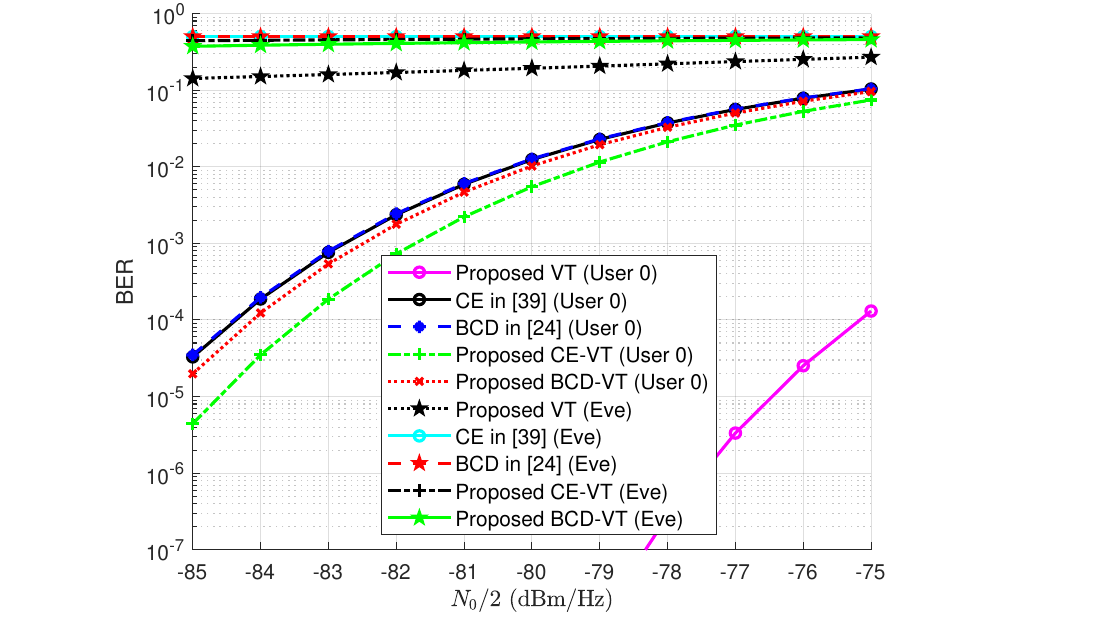}\\
	\caption{The BER of different methods for the zeroth user and Eve.}\label{fig:S4}
\end{figure}

\begin{figure}
	\centering
	\includegraphics[width=0.45\textwidth, trim = 50 260 20 280,clip]{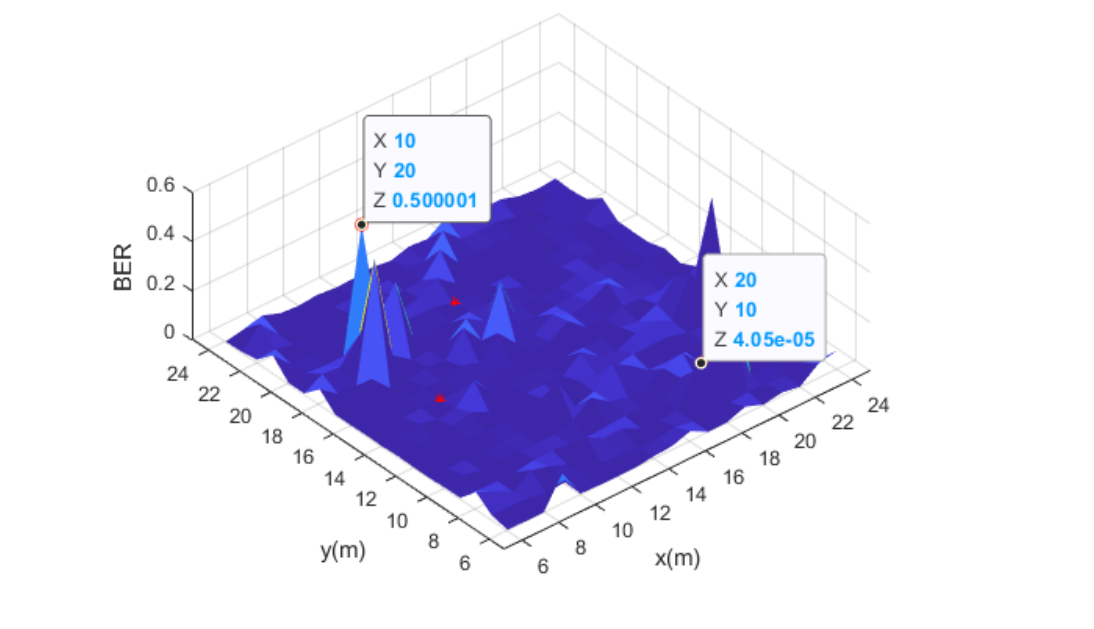}\\
	\caption{BER at different locations. The asterisk represents the user, and the circle is the Eve.}\label{fig:S2}
\end{figure}

Fig. \ref{fig:S4} illustrates the bit error rate (BER) of the zeroth user at different power spectral densities $N_0/2$. The proposed method has a low bit error rate (BER) at the user's location and a high BER at the Eve's location. If $N_0/2$ is less than $10^{-4.5}$, the VT method is used to maintain an extremely low BER level at the user's location, and the BER at the Eve's location gradually increases. If $N_0/2$ is less than $10^{-5.1}$, and the CE, BCD, CE-VT, and BCD-VT algorithms are used to maintain extremely low BER levels at the user's location. The BER of the CE and BCD methods at the Eve's location remains around 0.5, and the BER at the Eve's location gradually increases under the CE-VT and BCD-VT algorithms. At a signal-to-noise ratio (SNR) of $12dB$, the BER obtained using the BCD method is shown in Fig. \ref{fig:S2}.

\begin{figure}[h]
 	\centering
 	\includegraphics[width=0.45\textwidth, trim = 50 260 20 280,clip]{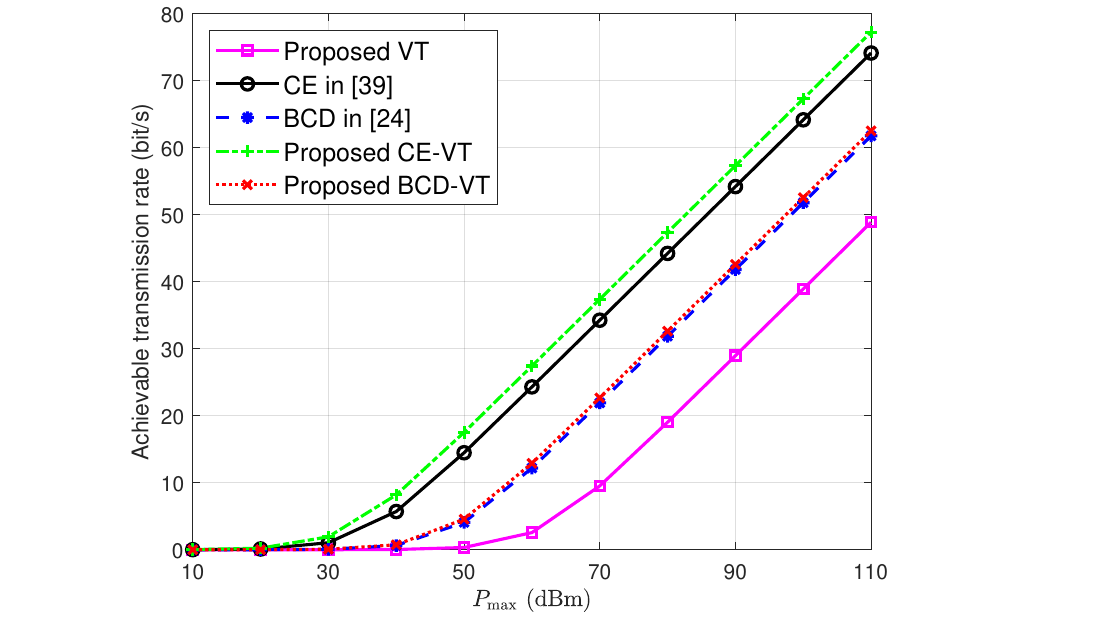}\\
 	\caption{The achievable transmission rate without uncertain component interference.}\label{fig:S6}
 \end{figure}

\begin{figure}[h]
 	\centering
 	\includegraphics[width=0.45\textwidth, trim = 50 260 20 280,clip]{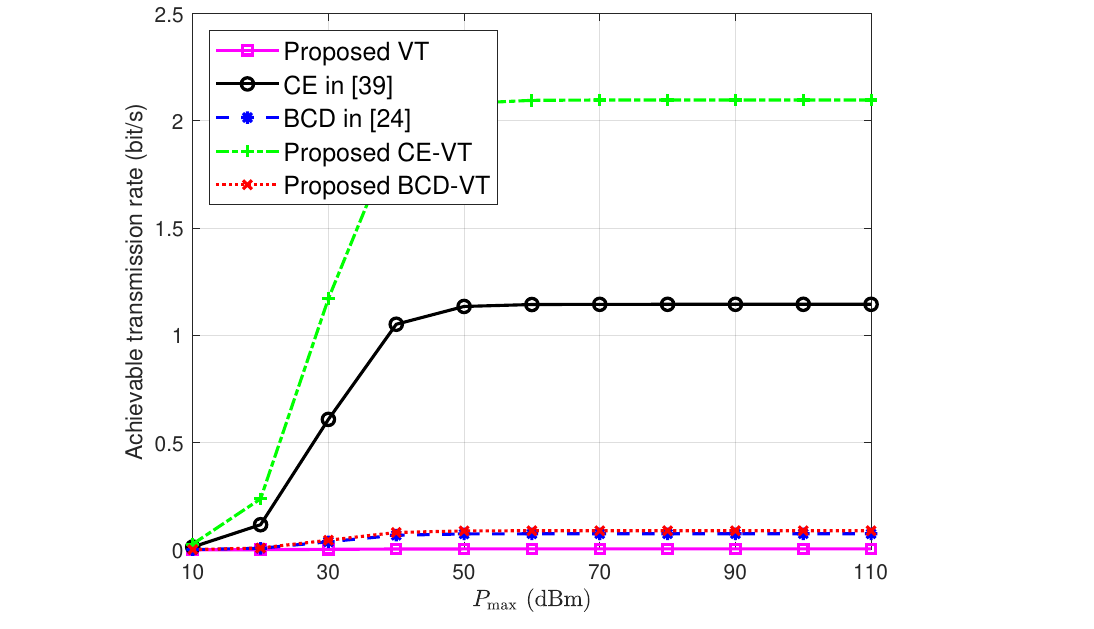}\\
 	\caption{The achievable transmission rate with uncertain component interference.}\label{fig:S6-1}
 \end{figure}

\begin{figure}[h]
 	\centering
 	\includegraphics[width=0.45\textwidth, trim = 50 260 20 280,clip]{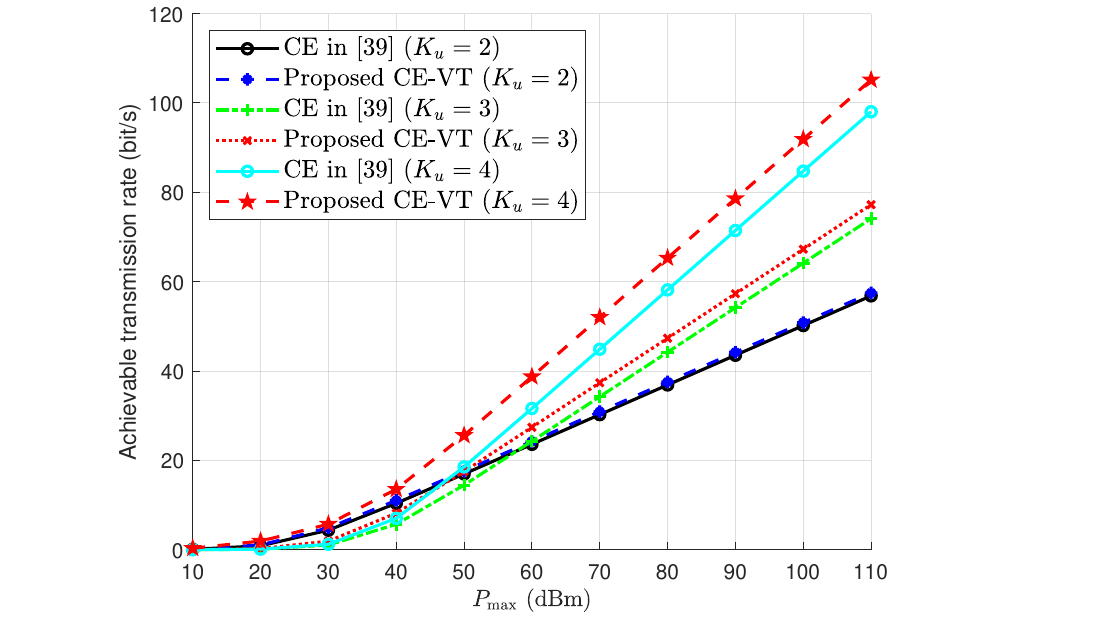}\\
 	\caption{The achievable transmission rate for different numbers of users.}\label{fig:S7}
 \end{figure}

Figs. \ref{fig:S6} and \ref{fig:S6-1} illustrate the achievable transmission rates with signal uncertain component interference and without uncertain component, respectively. In Fig. \ref{fig:S6}, as the maximum transmission power increases, the achievable transmission rate is increased. The achievable transmission rates with CE-VT and BCD-VT are superior to the CE method and BCD method, respectively. In Fig. \ref{fig:S6-1}, with uncertain component, the achievable transmission rates under BCD method and BCD-VT are relatively low, the CE method and CE-VT algorithm can achieve higher achievable transmission rates, which tend to stabilize at a maximum transmission power of $50dBm$. For the CE-VT algorithm, the rate improvement obtained is approximately double that of the CE method. Fig. $\ref{fig:S7}$ shows the achievable transmission rates for different numbers of users, demonstrating that the proposed scheme is effective for different numbers of users.


\section{Conclusion}\label{6}
A symbol level design scheme based on DM is proposed for the IRS-assisted UAV communication scenario. In order to realize the maximum transmission rate of UAV-DM system under the constraints of receiver symbol $e$, receiver sensitivity, maximum transmit power, IRS discrete phase shift, constant modulus and position range, the joint design of digital weight vector, UAV position and IRS phase shift matrix is studied. The scheme first obtains the sub-optimal UAV position by minimizing power. According to the obtained UAV position, the digital weight vector is scaled up to make full use of the given transmitting power. Then, based on the obtained UAV position and digital weight vector, a vector trajectory method is proposed to optimize the IRS phase shift matrix. This method has low complexity, and can be combined with the traditional discrete phase shift optimization method to obtain a higher transmission rate. Simulation results show that the proposed schemes can achieve low bit error rate at the user's location and high bit error rate at the eavesdropper's location, and has better transmission rate performance than the traditional CE and BCD methods. Particularly, the rate of the proposed CE-VT is approximately twice that of the CE and far beter than those of the remaining methods.


\renewcommand\refname{References}
\bibliographystyle{IEEEtran}
\bibliography{mybib}

%
%
%
%
%
%
\begin{IEEEbiographynophoto}{Maolin Li}
is currently pursuing the Ph.D. degree with the School of Information and Communication Engineer, Hainan University, China. His research interests include physical layer security and intelligent reflecting surface.
\end{IEEEbiographynophoto}
\begin{IEEEbiographynophoto}{Wei Gao}
received the B.E. degree in communication engineering and the Ph.D. degree of information and communication engineering from the Huazhong University of Science and Technology (HUST) in 2014 and 2020, respectively. He is currently a postdoc with Hainan University. His research interests include network architecture, wireless network access and radio resources allocation.
\end{IEEEbiographynophoto}
\begin{IEEEbiographynophoto}{Qi Wu}
received the Ph.D. degree from Southeast University, Nanjing, China, in 2009. He is currently an Associate Professor of Control Science and Engineering with the School of Electronic, Information and Electrical Engineering, Shanghai Jiao Tong University, Shanghai, China. His current research interests include pattern recognition and fault diagnosis.
\end{IEEEbiographynophoto}
\begin{IEEEbiographynophoto}{Feng Shu}
(Member, IEEE) was born in 1973. He received the B.S. degree from Fuyang Teaching College, Fuyang, China, in 1994, the M.S. degree from Xidian University, Xi’an, China, in 1997, and the
Ph.D. degree from Southeast University, Nanjing, China, in 2002. From September 2009 to September
2010, he was a Visiting Postdoctoral Fellow with the University of Texas at Dallas, Richardson, TX,
USA. From July 2007 to September 2007, he was a Visiting Scholar with the Royal Melbourne Institute
of Technology (RMIT University), Australia. From October 2005 to November 2020, he was with the School of Electronic and Optical Engineering, Nanjing University of Science and Technology, Nanjing, where he was promoted from an Associate Professor to a Full Professor of supervising Ph.D. students in 2013. Since December 2020, he has been with the School of Information and Communication Engineering, Hainan University, Haikou, China, where he is currently a Professor and a Supervisor of Ph.D. and graduate students. His research interests include wireless networks, wireless location, and array signal processing. He is awarded with the Leading-Talent Plan of Hainan Province in 2020, the Fujian Hundred-Talent Plan of Fujian Province in 2018, and the Mingjian Scholar Chair Professor in 2015. He has authored or coauthored more than 300 in archival journals with more than 150 papers on IEEE journals and 220 SCIindexed papers. His citations are more than 6000 times. He holds more than 40 Chinese patents and is also a PI or CoPI for seven national projects. He was an IEEE Transactions on Communications Exemplary Reviewer for 2020. He is currently an Editor of IEEE Wireless Communications Letters. He was an Editor of the IEEE Systems Journal from 2019 to 2021 and IEEE Access from 2016 to 2018.
\end{IEEEbiographynophoto}
\begin{IEEEbiographynophoto}{Cunhua Pan}
(Member, IEEE) received the B.S. and Ph.D. degrees from the School of Information Science and Engineering, Southeast University, Nanjing, China, in 2010 and 2015, respectively. From 2015 to 2016, he was a Research Associate with the University of Kent, U.K. He held a post-doctoral position at the Queen Mary University of London, U.K., from 2016 to 2019, where he is currently a Lecturer (Assistant Professor). His research interests mainly include ultra-dense C-RAN, machine learning, UAV, the Internet of Things, and mobile edge computing. He serves as a TPC Member for numerous conferences, such as ICC and GLOBECOM, and the Student Travel Grant Chair for ICC 2019. He also serves as an Editor for IEEE ACCESS.
\end{IEEEbiographynophoto}
\begin{IEEEbiographynophoto}{Di Wu}
was born in 1991. He received the M.S. degree in information and communication engineering from Hainan University, Haikou, China, in 2018. He is currently pursuing the Ph.D. degree in control science and engineering with the Department of Automation, Shanghai Jiao Tong University, Shanghai, China. He is currently a Lecturer with the School of Information and Communication Engineering, Hainan University. His research interests primarily revolve around nonlinear control of aerial vehicles, motion planning, and localization and mapping.
\end{IEEEbiographynophoto}
\vfill

\end{document}